\documentclass[pra, onecolumn, showpacs, secnumarabic]{revtex4}
\usepackage[cp1250]{inputenc}
\usepackage{indentfirst}
\usepackage{graphicx}
\usepackage{amsmath}
\usepackage{bbm} % $\mathbbm{} \mathbbmss{} \mathbbmss{}$
\newcommand{\sinc}{\hbox{sinc}}

\newcommand{\e}{\hbox{e}}
\newcommand{\tens}{\mathbf}

\begin{document}
%\maketitle
\title{Noise reduction in 3D noncollinear parametric amplifier}
\author{Piotr Migda{\l}}
\email{migdal@fuw.edu.pl}
\affiliation{Institute of Experimental Physics, University of Warsaw, Ho\.{z}a 69, PL-00-681 Warsaw, Poland}
\author{Wojciech Wasilewski}
\affiliation{Institute of Experimental Physics, University of Warsaw, Ho\.{z}a 69, PL-00-681 Warsaw, Poland}

\begin{abstract}
We analytically find an approximate Bloch-Messiah reduction of a noncollinear parametric amplifier pumped with a focused monochromatic beam. We consider type I phase matching. The results are obtained using a perturbative expansion and scaled to a high gain regime. They allow a straightforward maximization of the signal gain and minimization of the parametric fluorescence noise. 
We find the fundamental mode of the amplifier, which is an elliptic Gaussian defining the optimal seed beam shape. We conclude that the output of the amplifier should be stripped of higher order modes, which are approximately Hermite-Gaussian beams. Alternatively, the pump waist can be adjusted such that the amount of noise produced in the higher order modes is minimized.

\end{abstract}


\maketitle

\section{Introduction}

Parametric down-conversion is an important and versatile source of light. Its applications span from amplification of laser pulses to creation of strictly quantum states in the form of entangled photon pairs and squeezed vacuum. As optical parametric amplifiers rely on nonresonant nonlinear interaction, they can be utilized to amplify extremely wide bandwidth pulses.  One of the most appealing applications of this phenomenon is the Optical Parametric Chirped Pulse Amplifier (OPCPA)  \cite{DubietisOC92,DubietisSTiQEIJo06} in which ultrashort pulses are stretched, amplified and then compressed again.
The gain medium is typically pumped by a nanosecond second harmonic pulse from a Q-switched laser. Such amplifiers can provide a few femtosecond long multi-terawatt pulses \cite{RossJOSAB02,TavellaOE06}.

One of the most important and fundamental problems with this type of amplifiers is the presence of spontaneous parametric fluorescence, which unavoidably accompanies useful gain and compromises the pulse contrast ratio. This issue is central in many applications and has been extensively studied \cite{ArisholmJOSAB99,TavellaAPBLaO05,TavellaNJoP06}. Although the numerical results provided in \cite{TavellaNJoP06} are very accurate, they are based on Langevin noise equations \cite{GattiPRA97} developed for a cavity parametric amplifier. Consequently, the noise parameter has to rely on an experimental calibration. Furthermore, their numerical character does not give direct guidelines for noise reduction.

In this paper, we develop a simple analytical model of a 3D Noncollinear Optical Parametric Amplifier (NOPA). We assume type-I noncollinear phase matching in a second-order nonlinear crystal pumped with a monochromatic Gaussian beam. As a consequence, we are able to calculate the intensity of parametric fluorescence for any given signal gain, as well as its spatial and spectral distributions. Moreover, we find the optimal seed mode and pump waist for which spontaneous fluorescence is minimal.

Our model is built around the Bloch-Messiah reduction \cite{BraunsteinBM} of an optical parametric amplifier operated without pump depletion. It is a theorem which allows us to find a set of orthogonal, characteristic modes (or eigenmodes) of the amplifier and their respective gains. 
In the case of monochromatic pumping, the modes are the input and output beam shapes of the signal and idler fields at any pair of frequencies matching up to the pump. The main feature of the reduction is that if the amplifier is seeded with a beam matching its characteristic input mode, then this beam is amplified by the associated gain factor and assumes the output characteristic shape on the other end of the amplifier. 
In addition, spontaneous parametric fluorescence is produced in all of the characteristic output modes and the number of photons scattered into each of those modes is equal to the gain of that mode. In particular, the immediate conclusion from the reduction is that the optimal signal to parametric noise ratio is obtained when the seed beam has the shape of the fundamental input mode. The output of the amplifier may be spatially filtered to reject as many of the higher other modes as possible.

The paper is organized as follows. Section 2 presents general theory of parametric down-conversion, with emphasis on the 3D crystal pumped by a monochromatic wave. We introduce the Bloch-Messiah reduction theorem and discuss its general consequences. In Section 3 we derivate the characteristic modes and their gains for two particular settings - noncollinear with no walk-off and noncollinear with a significant walk-off. Section 4 summarizes the results of the analytical reduction. In Section 5 we compare analytical results with precise numerical calculations and discuss the accuracy of the approximations assumed. Section 6 describes behavior of the parametric gain in the intense pumping regime, provides estimation of the fluorescence noise in practical situations and gives solutions for the noise reduction. Finally, Section 7 concludes the paper and gives insight into possible applications and extensions.

\section{Bloch-Messiah reduction}

We aim at investigating the properties of a 3D parametric amplifier pumped by a strong, monochromatic laser beam of frequency $\omega_p$ and diameter $2w_p$. We assume that the signal and the spontaneous fluorescence fields are always so weak that they cannot influence the evolution of the pump. Within this approximation, the nonlinear interaction with the pump field couples pairs of conjugate frequencies, commonly called signal $\omega_s$ and idler $\omega_i=\omega_p-\omega_s$. Let us note that any two adjacent frequencies $\omega_s\ne\omega_s'$ evolve completely independently thanks to narrowband pumping. Therefore, amplification of a broadband seed pulse can be calculated separately for each of its monochromatic components, which may enter the crystal at different angles \cite{DanieliusOL1996}. Throughout the paper, we will analyze the interaction of a particular pair of frequencies $\omega_s$ and $\omega_i$ with the pump.

We choose a coordinate system where the $z$ axis lies along the central $k$-vector of the pump $k_p$ and the crystal faces are perpendicular to this axis. Then, it is most convenient to follow the convention of describing the evolution of the system as a function of $z$. We expand the electric field of each of the interacting waves in the basis of monochromatic plane waves parameterized by frequency $\omega$ and perpendicular components of the wave vector $k_x$ and $k_y$
\begin{align}
	E(t,x,y,z)=i\int d\omega dk_x dk_y \sqrt{\frac{\hbar \omega}{4\pi \epsilon_0 c
	\sqrt{n(\omega)}}}e^{-i\omega t+ixk_x +iyk_y} A(\omega,k_x,k_y,z) + \hbox{c.c.},
\end{align} 
where $A(\omega,k_x,k_y,z)$ is the $z$-dependent frequency-space amplitude of the wave.
With the above normalization $|A(\omega,k_x,k_y,z)|^2$ is the photon density.

\subsection{1-D amplifier}
Before analyzing a bulk amplifier let us briefly recall the basic concepts for a waveguide amplifier which can be reduced into a 2-mode amplifier for a defined signal frequency. This simple system will help in the comprehension of a full-fledged 3-D amplifier. Later on, the Bloch-Messiah decomposition will allow us to simplify the 3-D amplifier to a set of independent 2-mode amplifiers.

The pump field undergoes only dispersive evolution and its amplitude along the $z$ axis is $A_p(z)=P e^{i k_{p,z} z}$, where $k_{p,z}$ is the pump wavenumber and $P$ is the initial amplitude. The evolution of the signal amplitude $A_s(z)$ and the idler amplitude $A_i(z)$ is described by the equations of motion \cite{BandPRA94}
\begin{align}
	\frac{\partial}{\partial z}A_s(z) &= i k_{s,z} A_s(z)+\chi P \e^{i k_{p,z} z} A_i^*(z)\nonumber\\
	\frac{\partial}{\partial z}A_i(z) &= i k_{i,z} A_i(z)+\chi P \e^{i k_{p,z} z} A_s^*(z).\label{eq:motion}
\end{align}
The wave vectors for the signal and the idler are $k_{s,z}$ and $k_{i,z}$, respectively. The first term on the right hand side is responsible for linear propagation. The second one describes the nonlinear interaction with the coupling constant $\chi$. 
 
For a waveguide which starts at $-L/2$ and ends at $L/2$ one can obtain the following input-output relations:
\begin{align}
	(\e^{i \delta_3}A_s^{out})&=\cosh(\xi)(\e^{i \delta_1}A_s^{in})+\sinh(\xi)(\e^{i \delta_2}A_i^{in})^*\nonumber\\
	(\e^{i \delta_4}A_i^{out})&=\cosh(\xi)(\e^{i \delta_2}A_i^{in})+\sinh(\xi)(\e^{i \delta_1}A_s^{in})^*\label{eq:squeeze2mode}.
\end{align}
The upper indices $in$ and $out$ denote positions $z=-L/2$ and $z=L/2$, respectively. Phase terms with $\delta_1$, $\delta_2$, $\delta_3$ and $\delta_4$ depend on wave vector matching and we skip their explicit form for brevity. The intensity gain of the amplifier equals $\cosh^2 \xi$, where the gain parameter $\xi$ is proportional to the pump amplitude $P$. In addition to the amplified signal, the interaction produces spontaneous fluorescence. Quantum-mechanical considerations \cite{Loundon} show that the average number of photons generated this way equals $\sinh^2 \xi$.
This system is called a 2-mode or nondegenerate parametric amplifier \cite{BoydNLO}.  If one needs to calculate the amplification of a realistic laser pulse, the pulse has to be decomposed into monochromatic waves via Fourier transform, amplified as in (\ref{eq:squeeze2mode}) and composed back.

\subsection{Bulk amplifier}

\begin{figure}[htbp]
	\centering
		\includegraphics[width=0.50\textwidth]{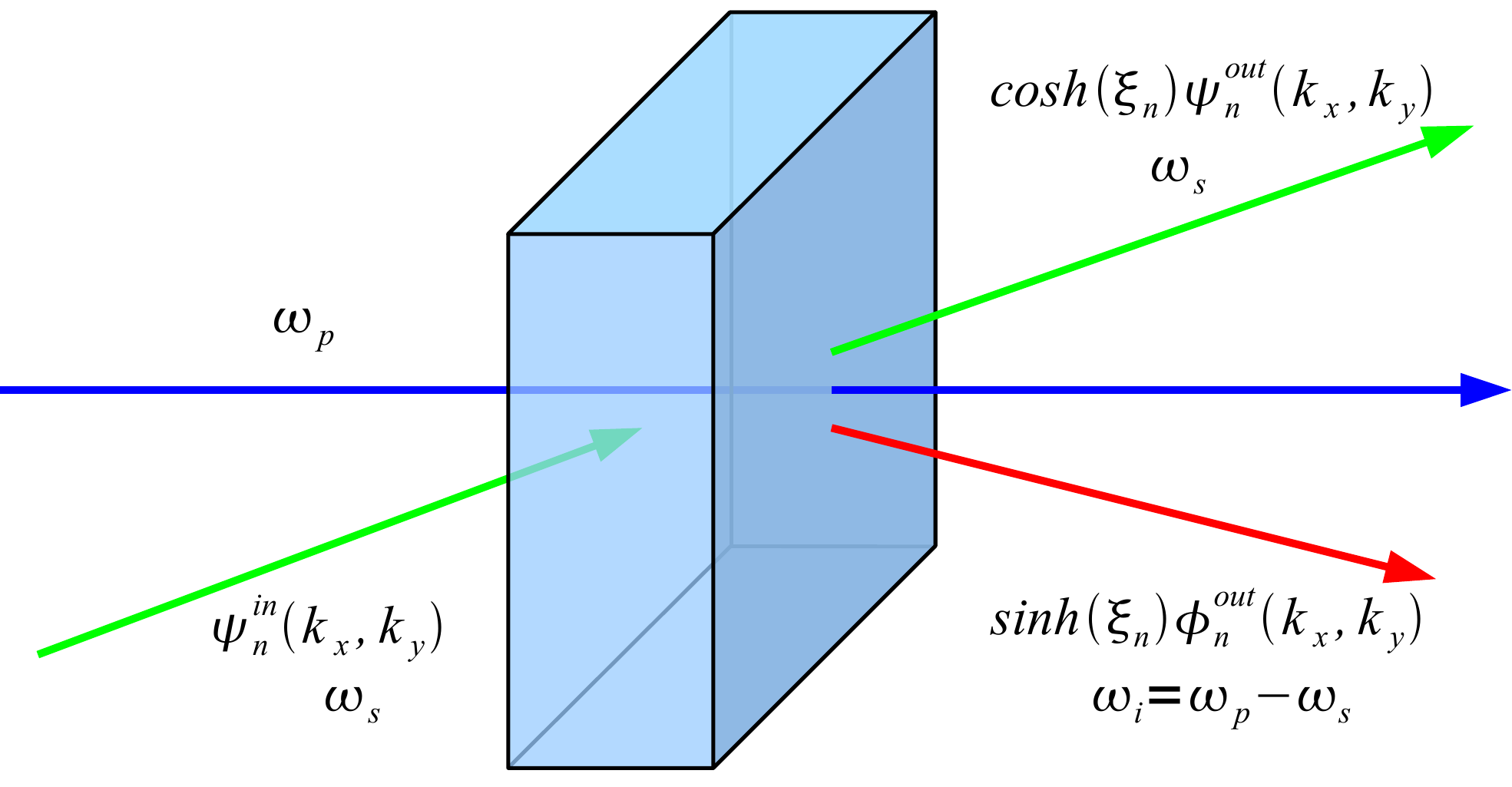}
	\caption{Schematic of a NOPA pumped with a focused monochromatic beam. When an incident signal beam is in the $\psi_n^{in}(\vec{k}_{s,\perp})$ mode, it produces the signal $\psi_n^{out}(\vec{k}_{s,\perp})$ at the output with the amplitude amplification of $\cosh(\xi_n)$ and the idler $\phi_n^{out}(\vec{k}_{i,\perp})$ with the amplitude $\sinh(\xi_n)$.}
	\label{fig:bloch-messiah}
\end{figure}

Let us now switch to a bulk crystal with the length $L$ in the $z$ direction and infinite in the $x$ and $y$ directions. The pump $A_p(\vec{k}_{p,\perp})$, the signal $A_s(\vec{k}_{s,\perp})$ and the idler $A_i(\vec{k}_{i,\perp})$ amplitudes have spatial freedom which is encoded in their dependence on the spatial wave vector $\vec{k}_{\perp}=(k_x,k_y)$. The infinite transversal size of the crystal allows interaction only of those signal and idler components for which transverse wave vectors sum up to the pump's. Therefore in each slice of the crystal with given $z$ only those interactions can take place, which preserve both the energy $\omega_p=\omega_s+\omega_i$ and the perpendicular components of the momentum
\begin{align}
	\vec{k}_{p,\perp}=\vec{k}_{s,\perp}+\vec{k}_{i,\perp}\label{eq:psiperp}.
\end{align}
Using these principles, an equation of motion for the waveguide amplifier \eqref{eq:motion} can be extended as follows
\begin{align}
	\frac{\partial}{\partial z}A_s(\vec{k}_{s,\perp},z)
	&= i k_{s,z} A_s(\vec{k}_{s,\perp},z)
	+ \chi \int d\vec{k}_{i,\perp}  A_p(\vec{k}_{s,\perp}+\vec{k}_{i,\perp}) \e^{i k_{p,z} z} A_i^*(\vec{k}_{i,\perp},z)\nonumber\\
	\frac{\partial}{\partial z}A_i(\vec{k}_{i,\perp},z)
	&= i k_{i,z} A_i(\vec{k}_{i,\perp},z)
	+ \chi \int d\vec{k}_{s,\perp} A_p(\vec{k}_{s,\perp}+\vec{k}_{i,\perp}) \e^{i k_{p,z} z} A_s^*(\vec{k}_{s,\perp},z)\label{eq:motion2d}.
\end{align}
As in the waveguide \eqref{eq:motion}, the first term on the right hand side represent the dispersive propagation. The second term is responsible for nonlinear interactions involving all the possible planewave components fulfilling \eqref{eq:psiperp}. 
The above equations are linear in the signal and idler amplitudes, and so are the input-output relations for the amplifier. Formally, they can be written down using input-output relations with integral kernels (or Green functions) $C_{ss}(\vec{k}_{s,\perp},\vec{k}_{s,\perp})$, $S_{si}(\vec{k}_{s,\perp},\vec{k}_{i,\perp})$, $C_{ii}(\vec{k}_{i,\perp},\vec{k}_{i,\perp})$, $S_{is}(\vec{k}_{i,\perp},\vec{k}_{s,\perp})$
\begin{align}
	A_s^{out}(\vec{k}_{s,\perp})&=
	\int d\vec{k}_{s,\perp}'
	 C_{ss}(\vec{k}_{s,\perp},\vec{k}_{s,\perp}') A_s^{in}(\vec{k}_{s,\perp}')
	+ \int d\vec{k}_{i,\perp} S_{si}(\vec{k}_{s,\perp},\vec{k}_{i,\perp}) A_i^{in*}(\vec{k}_{i,\perp})\nonumber\\
		A_i^{out}(\vec{k}_{i,\perp})&=
	\int d\vec{k}_{i,\perp}'
	 C_{ii}(\vec{k}_{i,\perp},\vec{k}_{i,\perp}') A_i^{in}(\vec{k}_{i,\perp}')
	+ \int d\vec{k}_{s,\perp} S_{is}(\vec{k}_{i,\perp},\vec{k}_{s,\perp}) A_s^{in*}(\vec{k}_{i,\perp})\label{eq:squeezespatialmodes}.
\end{align}
Since the integral kernels describe a reversible process, they turn out to be interdependent. The Bloch-Messiah theorem \cite{BraunsteinBM} relates their singular value decompositions (SVD): they have common modes and gain parameters, that is
\begin{align}
C_{ss}(\vec{k}_{s,\perp},\vec{k}_{s,\perp}')
&=\sum_{n=0}^\infty \psi_n^{out}(\vec{k}_{s,\perp}) \cosh(\xi_n) \psi_n^{in*}(\vec{k}_{s,\perp}'), & %\nonumber\\
C_{ii}(\vec{k}_{i,\perp},\vec{k}_{i,\perp}')
&=\sum_{n=0}^\infty \phi_n^{out}(\vec{k}_{i,\perp}) \cosh(\xi_n) \phi_n^{in*}(\vec{k}_{i,\perp}'),\nonumber\\
S_{si}(\vec{k}_{s,\perp},\vec{k}_{i,\perp})
&=\sum_{n=0}^\infty \psi_n^{out}(\vec{k}_{s,\perp}) \sinh(\xi_n) \phi_n^{in}(\vec{k}_{i,\perp}),& %\nonumber\\
S_{is}(\vec{k}_{i,\perp},\vec{k}_{s,\perp})
&=\sum_{n=0}^\infty \phi_n^{out}(\vec{k}_{i,\perp}) \sinh(\xi_n) \psi_n^{in}(\vec{k}_{s,\perp}).\label{eq:bloch-messiah-cs}
\end{align}
Above, $\psi_n^{in}(\vec{k}_{s,\perp})$, $\psi_n^{out}(\vec{k}_{s,\perp})$, $\phi_n^{in}(\vec{k}_{i,\perp})$ and $\phi_n^{out}(\vec{k}_{i,\perp})$ represent the input signal modes, the output signal modes, the input idler modes and the output idler modes, respectively, while $\xi_n$ are their gain parameters. Mathematically, $\psi_n^{in}(\vec{k}_{s,\perp})$, $\psi_n^{out}(\vec{k}_{s,\perp})$, $\phi_n^{in}(\vec{k}_{i,\perp})$ and $\phi_n^{out}(\vec{k}_{i,\perp})$ are four different sets of orthonormal functions. Physically, they represent characteristic beam shapes in the far field. Modes with different $n$ indices do not couple. More precisely, the input-output relations given by~\eqref{eq:squeezespatialmodes} can be brought to the canonical Bloch-Messiah form if we decompose the signal and the idler amplitudes in the bases of their characteristic modes
\begin{align}
	A_{s}^{in}(\vec{k}_{s,\perp}) & = \sum_{n=0}^\infty A_{s,n}^{in} \psi_n^{in}(\vec{k}_{s,\perp}), &
	A_{s}^{out}(\vec{k}_{s,\perp}) & = \sum_{n=0}^\infty A_{s,n}^{out} \psi_n^{out}(\vec{k}_{s,\perp}),\nonumber\\
	A_{i}^{in}(\vec{k}_{i,\perp}) & = \sum_{n=0}^\infty A_{i,n}^{in} \phi_n^{in}(\vec{k}_{i,\perp}), &
	A_{i}^{out}(\vec{k}_{i,\perp}) & = \sum_{n=0}^\infty A_{i,n}^{out} \phi_n^{out}(\vec{k}_{i,\perp}).\label{eq:ansi}
\end{align}
Substituting the above into \eqref{eq:squeezespatialmodes} and using \eqref{eq:bloch-messiah-cs} we find that the whole OPA can be decomposed into a set of independent 2-mode amplifiers
\begin{align}
	A_{s,n}^{out}&=\cosh(\xi_n)A_{s,n}^{in}+\sinh(\xi_n)A_{i,n}^{in*}\nonumber\\
	A_{i,n}^{out}&=\cosh(\xi_n)A_{i,n}^{in}+\sinh(\xi_n)A_{s,n}^{in*},\label{eq:squeeze2modes}
\end{align}
where $n$ indexes subsequent independent modes, conventionally sorted by their gains.
Note that the above equations have the same form as the input-output relation for the waveguide 2-mode amplifier given in \eqref{eq:squeeze2mode}. Similarly, the intensity gain of the $n$-th mode equals $\cosh^2 \xi_n$ and the average number of the photons spontaneously scattered into that mode equals $\sinh^2 \xi_n$. Therefore, we are interested in finding the characteristic beam shapes $\psi_n^{in}(\vec{k}_{s,\perp})$, $\psi_n^{out}(\vec{k}_{s,\perp})$ and gain parameters $\xi_n$, as this will provide the complete information. In particular, seeding the amplifier with its fundamental mode $\psi_0^{in}(\vec{k}_{s,\perp})$ would give us an advantage of obtaining maximal possible gain  $\cosh^2 \xi_0$. The idea of the decomposition is illustrated in Fig. \ref{fig:bloch-messiah}.

Let us make a nontrivial observation. For a special case of the crystal surfaces equally separated from the $z=0$ plane and with the real pump amplitude in $z=0$, we get a symmetry $S_{si}(\vec{k}_{s,\perp},\vec{k}_{i,\perp})=S_{is}^*(\vec{k}_{i,\perp},\vec{k}_{s,\perp})$ \cite{WasilewskiPRA06a}. When every gain parameter is unique $\xi_n\ne\xi_m$, the input and output signal modes are mirror reflections of each other, as are the input and output idler modes 
\begin{align}
	\psi_n^{out}(\vec{k}_{s,\perp})&=\psi_n^{in*}(\vec{k}_{s,\perp}), &
	\phi_n^{out}(\vec{k}_{i,\perp})&=\phi_n^{in*}(\vec{k}_{i,\perp}).\label{eq:symmetry}
\end{align}
This observation significantly simplifies the Bloch-Messiah reduction given in \eqref{eq:bloch-messiah-cs}. In particular it allows us to deduce all the characteristic mode functions from the SVD of $S_{is}(\vec{k}_{i,\perp},\vec{k}_{s,\perp})$.

Even though in and out modes are similar \eqref{eq:symmetry}, the behavior inside the crystal may be complex. Our description says nothing about what happens between the surfaces - we have only the input-output relations. The shape of the modes and their amplification vary in a number of parameters, including the crystal length $L$.

\subsection{Perturbative expansion}

In general the integral kernels defined in (\ref{eq:squeezespatialmodes}) cannot be found analytically. However, one can obtain closed form expressions in a low gain regime $\chi\rightarrow 0$. This is accomplished in two steps. First we assume zero nonlinearity $\chi=0$. The \eqref{eq:motion2d} can be immediately integrated as follows
\begin{align}
	A_s(\vec{k}_{s,\perp},z)&=A_s(\vec{k}_{s,\perp},z=0)\exp[i k_{s,z}(\vec{k}_{s,\perp})z], &
	A_i(\vec{k}_{i,\perp},z)&=A_i(\vec{k}_{i,\perp},z=0)\exp[i k_{i,z}(\vec{k}_{i,\perp})z],\label{eq:freepropagation}
\end{align}
which describes the linear propagation. To obtain the first-order approximation with respect to $\chi$, we need to integrate \eqref{eq:motion2d} over $-L/2<z<L/2$ with the zero-order approximation (\ref{eq:freepropagation}) substituted on the right hand side. Comparing the result to the definition of the integral kernels \eqref{eq:squeezespatialmodes}, we obtain
\begin{align}
	C_{ss}(\vec{k}_{s,\perp},\vec{k}_{s,\perp}) &=\exp(i k_{s,z}z)\delta(k_{s,z}'-k_{s,z}), & 
	C_{ii}(\vec{k}_{i,\perp},\vec{k}_{i,\perp}) &=\exp(i k_{i,z}z)\delta(k_{i,z}'-k_{i,z}).
\end{align}
In the first-order approximation the above functions describe only linear propagation. 
However, the remaining two kernels
\begin{align}
S_{si}(\vec{k}_{s,\perp},\vec{k}_{i,\perp})&=\chi L \e^{-i (k_{i,z}-k_{s,z}) L/2} A_p(\vec{k}_{i,\perp}+\vec{k}_{s,\perp})\sinc\left( \Delta k L/2 \right)\nonumber\\
S_{is}(\vec{k}_{i,\perp},\vec{k}_{s,\perp})&=\chi L \e^{i (k_{i,z}-k_{s,z}) L/2} A_p(\vec{k}_{i,\perp}+\vec{k}_{s,\perp})\sinc\left( \Delta k L/2 \right),\label{eq:pert_sis}
\end{align}
describe down-conversion and give us insight into the characteristic modes of the Bloch-Messiah reduction \eqref{eq:bloch-messiah-cs}. We  call $S_{is}$ the scattering kernel, as it describes amplitudes with which the incident signal creates the 'scattered' idler beam. The $\Delta k$ %, which implicitly depends on $\vec{k}_{i,\perp}$ and $\vec{k}_{s,\perp}$ 
is the wave vector mismatch along the $z$ axis in this process
\begin{align}
	\Delta k=k_{s,z}+k_{i,z}-k_{p,z}\label{eq:dkz}.
\end{align}
The $z$ components of the wave vectors are functions of their transverse wave vector $k_{s,z}=k_{s,z}(\vec{k}_{s,\perp})$, $k_{i,z}=k_{i,z}(\vec{k}_{i,\perp})$ and $k_{p,z}=k_{p,z}(\vec{k}_{i,\perp}+\vec{k}_{s,\perp})$. %In particular, the mismatch $\Delta k$ is a function of $\vec{k}_{s,\perp}$ and $\vec{k}_{i,\perp}$.

Once we perform SVD of the scattering kernel $S_{is}$ \eqref{eq:pert_sis} we find all the characteristic modes and gain parameters of the Bloch-Messiah reduction \eqref{eq:bloch-messiah-cs} with the help of the identity \eqref{eq:symmetry}. Obtaining an approximate analytical SVD of the scattering kernel $S_{is}$ is described in the next section. It requires several nontrivial steps. However, we will always separate the variables in cylindrical coordinates as follows
\begin{align}
	S_{is}(k_{i,\perp},\varphi_i;k_{s,\perp},\varphi_s)&=K(k_{i,\perp},k_{s,\perp}) \Phi(\varphi_i,\varphi_s)\nonumber\\
		&=	\left[ \sum_n \phi_n^{K,out}(k_{i,\perp})\cdot \xi_n^K\cdot \psi_n^{K,in}(k_{s,\perp}) \right]
		\left[ \sum_m  \phi_m^{\Phi,out}(\varphi_i)\cdot \xi_m^\Phi\cdot \psi_m^{\Phi,in}(\varphi_s) \right]\nonumber\\
	&=	\sum_n \sum_m
	 \underbrace{\left[\phi_n^{K,out}(k_{i,\perp}) \phi_m^{\Phi,out}(\varphi_i)\right]}_{\phi_{n,m}^{out}(k_{i,\perp},\varphi_i)}
	 \underbrace{\xi_n^K \xi_m^\Phi}_{\xi_{n,m}}
	  \underbrace{\left[\psi_n^{K,in}(k_{s,\perp}) \psi_m^{\Phi,in}(\varphi_s)\right]}_{\psi_{n,m}^{out}(k_{s,\perp},\varphi_s)},
	\label{eq:svd_separation}
\end{align}
where functions and singular values with upper indices $K$ and $\Phi$ are singular value decompositions of  $K(k_{i,\perp},k_{s,\perp})$ and $\Phi(\varphi_i,\varphi_s)$, respectively.
Above we express the SVD of $S_{is}$ as a product of SVDs of its radial and angular parts, $K(k_{i,\perp},k_{s,\perp})$ and $\Phi(\varphi_i,\varphi_s)$. 
Note that this is also an approximation which relies on the fact that the radial mode functions $\phi_n^{K,out}(k_{i,\perp})$ and $\psi_n^{K,in}(k_{s,\perp})$ are centered around certain $k_\perp \simeq k_0$ and do not extend towards the origin $k_\perp \simeq 0$.

\section{Analytical approximation}
\begin{figure}[htbp]
	\centering
		\includegraphics[width=1.0\textwidth]{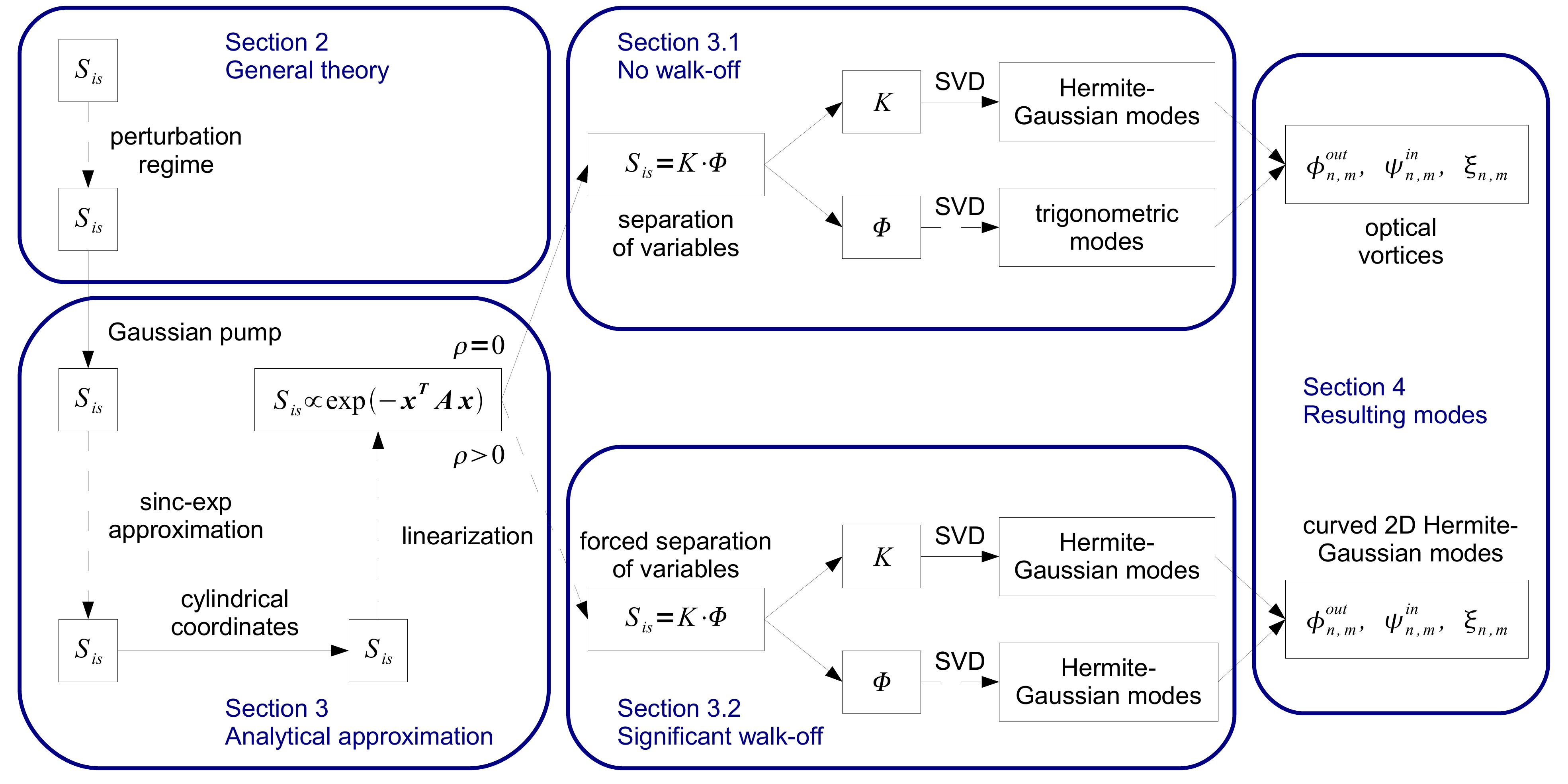}
	\caption{Schematic of the procedures applied in Section 3 along with parts of Section 2 and Section 4. Solid lines stand for equalities, while dashed lines for approximations. Final results are summarized and discussed in Section 4.}
	\label{fig:overview}
\end{figure}

Our task is to find the SVD of the scattering kernel $S_{is}(\vec{k}_{i,\perp},\vec{k}_{s,\perp})$ given in \eqref{eq:pert_sis} for a Gaussian pump beam. We will accurately find a mode with the highest amplification. To obtain the result we go through several approximations as illustrated in Fig.~\ref{fig:overview}. They include replacing sinc with a Gaussian, Taylor expansion of  the phase mismatch $\Delta k L/2$ and the separation of variables. We will consider two different physical settings - noncollinear without pump walk-off and noncollinear with a significant pump walk-off. For each of them, we obtain closed form Bloch-Messiah reduction, as summarized in Section 4. We restrict ourselves to type I down-conversion, although we conjecture that similar approach may work also for type II process.

We choose Gaussian pump beam profile with waist $w_p$ and the wavefront parallel to the crystal surface. The pump amplitude in spatial frequency domain is
\begin{align}		A_p(k_{p,x},k_{p,y})=P \frac{w_p}{\sqrt{\pi}}\exp\left[-\frac{w_p^2}{2}(k_{p,x}^2+k_{p,y}^2)\right].\label{eq:pump}
\end{align}
where $P$ is an amplitude factor.

\begin{figure}[htbp]
	\centering
		\includegraphics[width=0.50\textwidth]{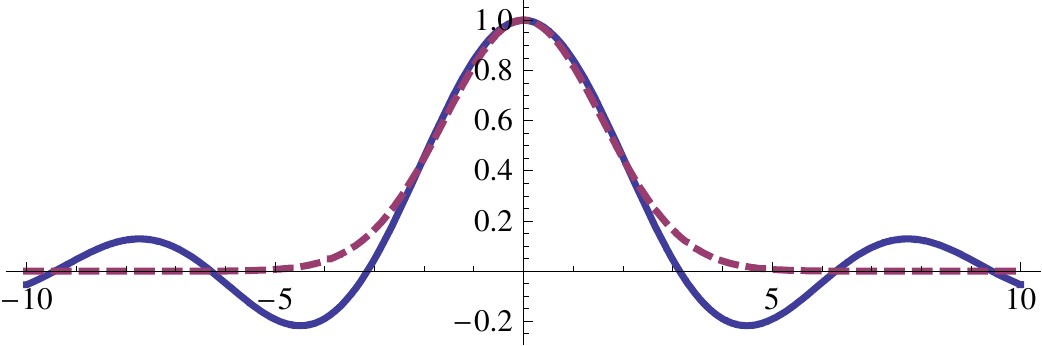}
	\caption{Plot of $\sinc(x)$ (solid line) and $\exp(-x^2/5)$ (dashed line). In the approximation of sinc with $\exp(-x^2/\kappa)$, reasonable values of the Gaussian width coefficient $\kappa$ spread from $4.16$ to $6.28$. One may use the closest Gaussian in the norm distance ($\kappa=4.16$), set the same FWHM ($\kappa=5.18$), compare Taylor series up to the second order ($\kappa=6$) or guarantee the same volume ($\kappa=6.28$). In various papers \cite{ChwedenczukPRA08b,WasilewskiPRA06a,DraganPRA04,KolenderskiPRA09,Uren03} different $\kappa$ values were used. We arbitrarily adopt $\kappa=5$.}
	\label{fig:sinc}
\end{figure}

In the first step we approximate $\sinc(\Delta k / L)$ with a Gaussian. This erases the oscillating tails but enables further calculations \cite{KolenderskiPRA09}.  We  write
\begin{align}
	\sinc \left(\frac{\Delta k L}{2}\right) \approx \exp \left[-\frac{(\Delta k L)^2}{20}\right]\label{eq:sinc5},
\end{align}
bearing in mind we chose arbitrarily the width of Gaussian fit as illustrated in Fig. \ref{fig:sinc}. Note that the length of the crystal $L$ appears only in the argument of $\sinc$ in (\ref{eq:pert_sis}). Consequently, the uncertainty in the numerical factor in the above approximation can be understood as a $\approx10\%$ uncertainty of the crystal length $L$.

Further steps involve Gaussian approximation of the scattering kernel. The task is simple but not straightforward. We need to choose the suitable variables as well as the right Taylor series approximation.

%========================
\begin{figure}[htbp]
	\centering
		\includegraphics[width=0.50\textwidth]{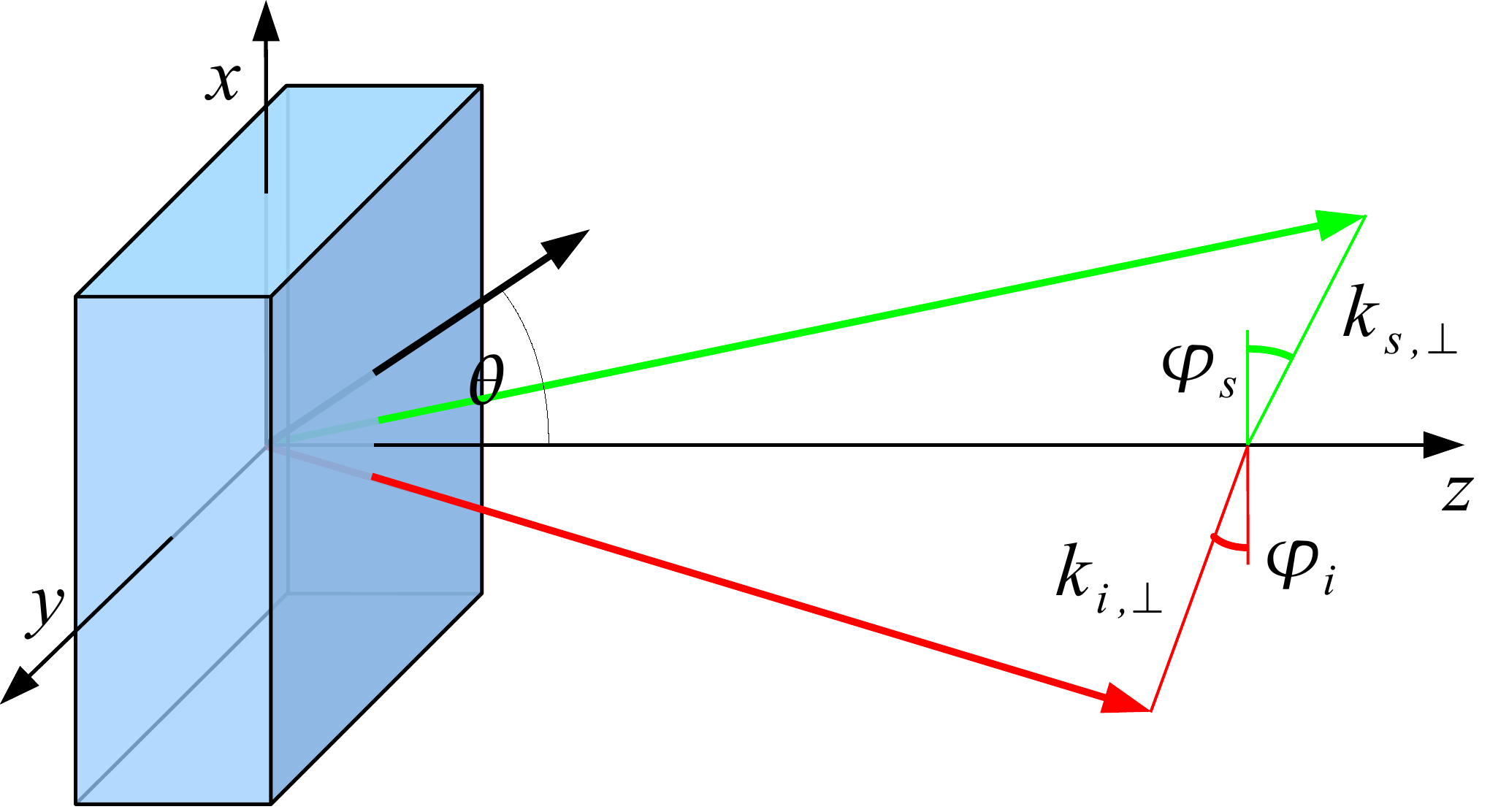}
	\caption{The crystal in a coordinate system. In both $x$ and $y$ direction the crystal is infinite, while in $z$ its length is $L$. The optical axis lies in the $xz$ plane, at the angle $\theta$ to the $z$ axis.}
	\label{fig:xyz_cone}
\end{figure}

Since the amplifier discussed has nearly cylindrical symmetry, it is convenient to introduce cylindrical coordinates in the following form
\begin{align}
\vec k_{s,\perp}&= \left[k_{s,\perp}\cos(\varphi_s),k_{s,\perp}\sin(\varphi_s)\right] &
\vec k_{i,\perp}&= 
\left[-k_{i,\perp}\cos(\varphi_i),-k_{i,\perp}\sin(\varphi_i)\right] \label{eq:cylindrical}.
%	k_{s,x}&=k_{s,\perp} \cos(\varphi_s)\nonumber\\
%	k_{s,y}&=k_{s,\perp} \sin(\varphi_s)\nonumber\\
%	k_{i,x}&=- k_{i,\perp} \cos(\varphi_i)\nonumber\\
%	k_{i,y}&=- k_{i,\perp} \sin(\varphi_i)
\end{align}
Note that angles on the parametric down-conversion cone $\varphi_s$ and $\varphi_i$ are measured from $0^\circ$ and $180^\circ$, respectively (Fig. \ref{fig:xyz_cone}). Then the phase matching is achieved for $\varphi_s\approx\varphi_i$. In cylindrical coordinates with approximation \eqref{eq:sinc5} the scattering kernel (\ref{eq:pert_sis}) reads
\begin{align}
	S_{is}(k_{i,\perp},\varphi_i;k_{s,\perp},\varphi_s)=\chi P L \frac{w_p}{\sqrt{\pi}} \e^{i (k_{i,z}-k_{s,z}) L/2} \exp\left[-\frac{k_{s,\perp}^2+k_{i,\perp}^2-2k_{s,\perp} k_{i,\perp} \cos(\varphi_s-\varphi_i)}{2/w_p^2}-\frac{1}{20}\Delta k^2 L^2\right]\label{eq:kernelexp}.
\end{align}
The scattering kernel achieves its maximal absolute value when $\vec{k}_{s,\perp}+\vec{k}_{i,\perp}=0$ and there is no mismatch $\Delta k = 0$. We will linearize $\Delta k$ around this point, which physically corresponds to the cone of ideal phase matching.

%As we are going to linearize the exponent in \eqref{eq:kernelexp}, we naturally apply the Taylor series approximation around the ideal down-conversion cone.  

\begin{figure}[htbp]
	\centering
		\includegraphics[width=0.50\textwidth]{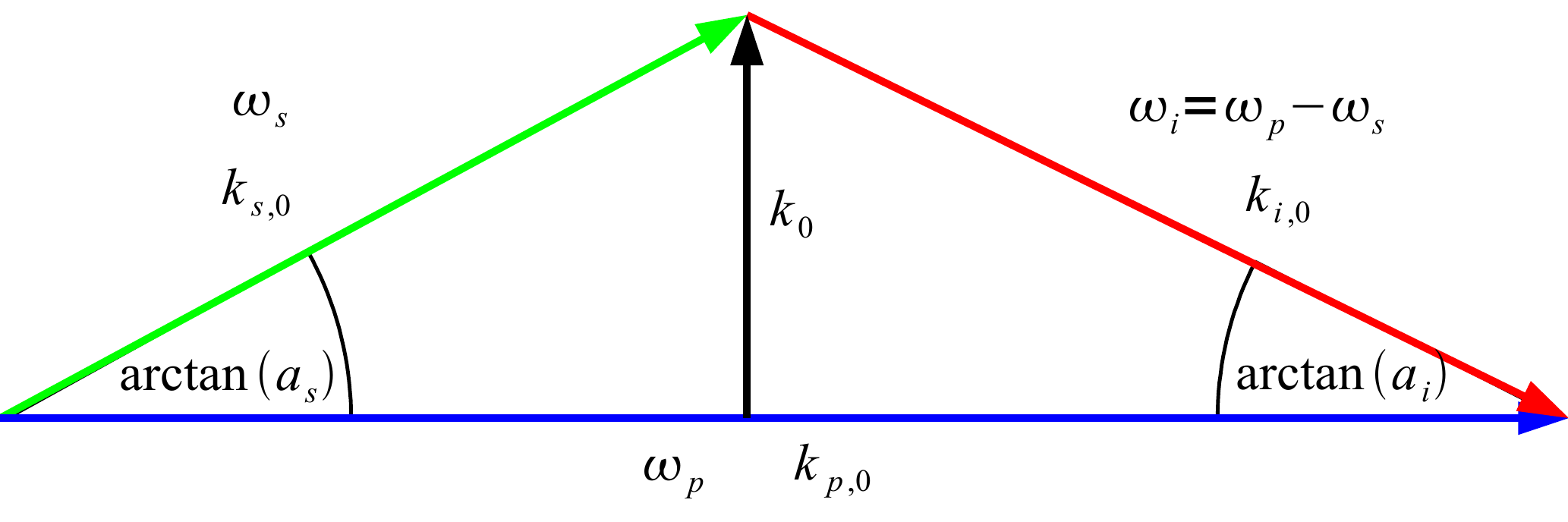}
	\caption{Schematic of ideal phase matching. The signal and the idler interact with the pump central component. Their respective wavenumbers are $k_{s,0}$, $k_{i,0}$ and $k_{p,0}$. The perpendicular wave vector component of both the signal and the idler is $k_0$. Inside the crystal, the signal travels at the angle $\arctan(a_s)\approx a_s$ and the idler at $\arctan(a_i)\approx a_i$. }
	\label{fig:ideal_match}
\end{figure}

To calculate the phase mismatch $\Delta k$ given in \eqref{eq:dkz} we need expressions for the $z$ components of the wave vectors of the interacting waves. We consider type I down-conversion with the signal and the idler propagating as ordinary waves and the pump as extraordinary. The optical axis lies in the $xz$ plane, at the angle $\theta$ to the $z$ axis. We find that
\begin{align}
k_{s,z}&=\sqrt{\frac{\omega_s^2\ n_o^2(\omega_s)}{c^2}-k_{s,\perp}^2}\label{eq:ksipz}\\
k_{i,z}&=\sqrt{\frac{(\omega_p-\omega_s)^2 n_o^2(\omega_p-\omega_s)}{c^2}-k_{i,\perp}^2}\nonumber\\
k_{p,z}&=\left(k_{s,x}+k_{i,x}\right)
\frac{\sin(2\theta)\left[n^2_o(\omega_p)-n^2_e(\omega_p)\right]}{2\left[n_o^2(\omega_p)\sin^2(\theta)+n_e^2(\omega_p)\cos^2(\theta)\right]}
+\frac{n_o(\omega_p)n_e(\omega_p)}{n_o^2(\omega_p)\sin^2(\theta)+n_e^2(\omega_p)\cos^2(\theta)}\nonumber\\
&\times \sqrt{\left[\frac{\omega^2}{c^2}-\frac{(k_{s,y} +k_{i,y})^2}{n_o^2(\omega_p)}\right]
\left[n_o^2(\omega_p)\sin^2(\theta)+n_e^2(\omega_p)\cos^2(\theta)\right]-(k_{s,x}+k_{i,x})^2},\nonumber
\end{align} 
where $c$ is the speed of light, $n_o(\omega)$ and $n_e(\omega)$ are ordinary and extraordinary refractive indices, respectively. 
The lengths of the wave vectors of the interacting waves are  $k_{s,0}$, $k_{i,0}$ and $k_{p,0}$ and $\rho$ is the tangent of the pump walk-off as illustrated in Fig.~\ref{fig:ideal_match}
\begin{align}
	k_{s,0}&=\frac{\omega_s}{c} n_o(\omega_s) &
	k_{i,0}&=\frac{\omega_p-\omega_s}{c} n_o(\omega_p-\omega_s)\nonumber\\
	k_{p,0}&=\frac{\omega_p}{c} \frac{n_o(\omega_p)n_e(\omega_p)}{\sqrt{n_o^2(\omega_p)\sin^2(\theta)+n_e^2(\omega_p)\cos^2(\theta)}} &
	\rho&=\frac{\sin(2\theta)\left[n^2_o(\omega_p)-n^2_e(\omega_p)\right]}{2\left[n_o^2(\omega_p)\sin^2(\theta)+n_e^2(\omega_p)\cos^2(\theta)\right]}.\label{eq:ktot}
\end{align}

The phase mismatch vanishes $\Delta k=0$ when $\varphi_s=\varphi_i$ and the perpendicular wave vectors components of the signal and the idler are equal $k_{s,\perp}=k_{i,\perp}=k_0$ with
\begin{align}
	k_0&=\frac{1}{2 k_{p,0}}\sqrt{2(k_{s,0}^2k_{i,0}^2+k_{p,0}^2k_{s,0}^2+k_{p,0}^2k_{i,0}^2)
	-k_{s,0}^4-k_{i,0}^4-k_{p,0}^4}.\label{eq:k0}%\\
%	&=\frac{1}{2 k_{p,0}}\sqrt{(k_{s,0}+k_{i,0}+k_{p,0})(-k_{s,0}+k_{i,0}+k_{p,0})(k_{s,0}-k_{i,0}+k_{p,0})(k_{s,0}+k_{i,0}-k_{p,0})}.
\end{align}
The linearization of the $\Delta k$ given in \eqref{eq:dkz} around its zero with respect to $k_{s,\perp}$, $k_{i,\perp}$ and $(\varphi_s-\varphi_i)$ using \eqref{eq:ksipz} gives us
\begin{align}
	\Delta k \approx \left[a_s-\rho \cos({\scriptstyle\frac{\varphi_s+\varphi_i}{2}})\right](k_{s,\perp}-k_0)+\left[a_i+\rho \cos({\scriptstyle\frac{\varphi_s+\varphi_i}{2}})\right](k_{i,\perp}-k_0)-k_0 \rho \sin({\scriptstyle\frac{\varphi_s+\varphi_i}{2}}) (\varphi_s-\varphi_i).\label{eq:Dkapprox}
\end{align}
Both $a_s={k_0}/{\sqrt{k_{s,0}^2-k_0^2}}$ and $a_i={k_0}/{\sqrt{k_{i,0}^2-k_0^2}}$  have an easy interpretation. Each of them is a tangent of the angle at which the signal or the idler travels inside the crystal in an ideal phase matching setting. Since $\rho$ \eqref{eq:ktot} is the tangent of the pump walk-off angle, the sum and difference of $a_s$ or $a_i$ and $\rho$ reflect how the signal or the idler diverge form the pump.

The Gaussian pump \eqref{eq:pump} may be approximated by Gaussian function also in cylindrical coordinates \eqref{eq:cylindrical}. For the exponent we obtain
\begin{align}
	k_{s,\perp}^2+k_{i,\perp}^2-2k_{s,\perp} k_{i,\perp} \cos(\varphi_s-\varphi_i)
	\approx k_0^2 (\varphi_s-\varphi_i)^2+(k_{s,\perp}-k_{i,\perp})^2,\label{eq:pumpapprox}
\end{align}
which is valid as long as $1/w_p\ll k_0$, that is the down-conversion cone opening angle is much bigger than the pump beam divergence.

Consequently, the scattering kernel $S_{is}$ given in \eqref{eq:kernelexp} substituted with \eqref{eq:Dkapprox} and \eqref{eq:pumpapprox} becomes a Gaussian function in $k_{s,\perp}$, $k_{i,\perp}$ and $(\varphi_s-\varphi_i)$, that is
\begin{align}
	S_{is}(k_{i,\perp},\varphi_i;k_{s,\perp},\varphi_s)=\chi P L\frac{w_p}{\sqrt{\pi}} \e^{i (k_{i,z}-k_{s,z}) L/2} \exp\left( - \textbf{x}^T \tens{A} \textbf{x} \right),\label{eq:kernelexpquadratic}
\end{align}
where $\textbf{x}=\left[(\varphi_s-\varphi_i),(k_{s,\perp}-k_0),(k_{i,\perp}-k_0)\right]$ is the deviation from ideal phase matching. The quadratic form coefficients are
\begin{align}
	\tens{A}&=\left[
	\begin{array}{ccc}
	A_{\varphi\varphi} & A_{\varphi s} & A_{\varphi i} \\
	A_{s \varphi} & A_{s s} & A_{s i} \\
	A_{i \varphi} & A_{i s} & A_{i i}
\end{array} \right]
	=\underbrace{\left[
	\begin{array}{ccc}
	\frac{k_0^2 w_p^2}{2}+\frac{L^2}{20}k_0^2 \rho^2 \sin^2({\scriptstyle\frac{\varphi_s+\varphi_i}{2}}) & 0 &	0 \\
	0 & 0 & 0 \\
	0 & 0 & 0
\end{array}	\right]}_{\hbox{the angular part}}\label{eq:quadratic}\\
	%===
	&+\underbrace{\left[
	\begin{array}{ccc}
	0 & 0 & 0 \\
	0 &
	 	 \frac{w_p^2}{2}+\frac{L^2}{20}\left[a_s-\rho \cos({\scriptstyle\frac{\varphi_s+\varphi_i}{2}})\right]^2	&
	 -\frac{w_p^2}{2}+\frac{L^2}{20}\left[a_s-\rho \cos({\scriptstyle\frac{\varphi_s+\varphi_i}{2}})\right]\left[a_i+\rho \cos({\scriptstyle\frac{\varphi_s+\varphi_i}{2}})\right]\\
	 0 & 
	 -\frac{w_p^2}{2}+\frac{L^2}{20}\left[a_s-\rho \cos({\scriptstyle\frac{\varphi_s+\varphi_i}{2}})\right]\left[a_i+\rho \cos({\scriptstyle\frac{\varphi_s+\varphi_i}{2}})\right] &
	 \frac{w_p^2}{2}+\frac{L^2}{20}\left[a_i+\rho \cos({\scriptstyle\frac{\varphi_s+\varphi_i}{2}})\right]^2
	\end{array}
	\right]}_{\hbox{the radial part}}\nonumber\\
%===
	&+\underbrace{\frac{L^2}{20} k_0 \rho \sin({\scriptstyle\frac{\varphi_s+\varphi_i}{2}})\left[
	\begin{array}{ccc}
	0 &
	 - \left[a_s-\rho \cos({\scriptstyle\frac{\varphi_s+\varphi_i}{2}})\right]&
	 - \left[a_i+\rho \cos({\scriptstyle\frac{\varphi_s+\varphi_i}{2}})\right] \\
	 - \left[a_s-\rho \cos({\scriptstyle\frac{\varphi_s+\varphi_i}{2}})\right] & 	 0	&	 0 \\
	 - \left[a_i+\rho \cos({\scriptstyle\frac{\varphi_s+\varphi_i}{2}})\right] & 	 0 &	 0
	\end{array}
	\right]}_{\hbox{the mixed part}}\nonumber
\end{align}
The matrix $\tens{A}$ is presented as a sum of different parts to facilitate the separation of variables. We call those components the angular part, the radial part and the mixed part. Once the mixed part is zero, the scattering kernel $S_{is}$ may be written as a product of two kernels, as outlined in (\ref{eq:svd_separation}). We will develop two different approximations: one for no pump walk-off $\rho=0$ and the other one for a significant pump walk-off. The aim of further steps is to neglect the mixed terms and get rid of trigonometric functions.

\subsection{No walk-off}
In some special situations there may be no walk-off ($\rho=0$). This is particularly easy to solve, as the quadratic form matrix (\ref{eq:kernelexpquadratic}) can be factorized into the angular and the radial part (as the mixed part vanishes)
\begin{align}
	\tens{A}=\left[
	\begin{array}{ccc}
	\frac{k_0^2 w_p^2}{2} & 0 & 0 \\
	 0 &
	 	 \frac{w_p^2}{2}+\frac{L^2}{20} a_s^2	&
	 -\frac{w_p^2}{2}+\frac{L^2}{20}a_s a_i \\
	 0 & 
	 -\frac{w_p^2}{2}+\frac{L^2}{20} a_s a_i &
	 \frac{w_p^2}{2}+\frac{L^2}{20}a_i^2
	\end{array}
	\right].\label{eq:quadratic_nowalk-off}
\end{align}
As the scattering kernel may be factorized $S_{is} = K(k_{i,\perp},k_{s,\perp}) \Phi(\varphi_i,\varphi_s)$ as in \eqref{eq:svd_separation}, we need to find the SVDs of two separate parts. To deal with the angular part we write
\begin{align}
	\Phi(\varphi_i,\varphi_s) = \exp\left[-\frac{k_0^2 w_p^2}{2}(\varphi_s-\varphi_i)^2\right]
	&=\frac{1}{\sqrt{2\pi}}\frac{1}{k_0 w_p}\int_{-\infty}^\infty dm \exp\left[-\frac{m^2}{2k_0^2 w_p^2}+im(\varphi_s-\varphi_i)\right]\nonumber\\
	&\approx \frac{1}{\sqrt{2\pi}}\frac{1}{k_0 w_p} \sum_{m=-\infty}^\infty
	\exp(-im\varphi_i)
	\exp\left(-\frac{m^2}{2k_0^2 w_p^2}\right)
	\exp(+im \varphi_s).\label{eq:angularexp}
\end{align}
We applied the Fourier representation of a Gaussian function and then we approximated the integration by summation. The second step required a moderately slowly changing function, which is guaranteed by $1/w_p\ll k_0$, already required in the previous step \eqref{eq:pumpapprox}.

The remaining radial part of the kernel $K(k_{i,\perp},k_{s,\perp})$ is a quadratic form of $(k_{s,\perp}-k_0)$ and $(k_{i,\perp}-k_0)$. It can be directly decomposed with the Mehler's Hermite polynomial formula \cite{BaroneAJP03}, which reads
\begin{align}
	\exp\left[-\frac{1+\mu^2}{2(1-\mu^2)}(x^2+y^2)+\frac{2 \mu x y}{1-\mu^2}\right]=
	\sqrt{\pi}\sqrt{1-\mu^2}\sum\limits_{n=0}^{\infty}\mu^n u_n(x)u_n(y),\nonumber\\
	u_n(x)=(2^n n!)^{-\frac{1}{2}}H_n(x)\exp(-x^2/2),\label{eq:hopropagator}
\end{align}
where $x$, $y$ and $\mu$ are real numbers, while $u_n(x)$ are the Hermite-Gaussian modes. Comparing the coefficient of the radial part $K(k_{i,\perp},k_{s,\perp})$ found in the lower-right part of \eqref{eq:quadratic_nowalk-off} with the general Mehler's formula, we get

\begin{align}
	K(k_{i,\perp},k_{s,\perp})&=\chi P L \frac{w_p}{\sqrt{\pi}}  \e^{i (k_{i,z}-k_{s,z}) L/2}\nonumber\\
	&\times\exp\left[ - A_{ss}(k_{s,\perp}-k_0)^2 - 2A_{si}(k_{s,\perp}-k_0)(k_{i,\perp}-k_0)- A_{ii}(k_{i,\perp}-k_0)^2 \right]\nonumber\\
	&=\sum_{n=0}^\infty \left\{ \e^{i k_{s,z} L/2} \frac{u_n\left[w_{i,r}(k_{i,\perp}-k_0)\right]}{\sqrt{1/w_{i,r}}} \right\}
	\chi P L \frac{w_p}{\sqrt{w_{s,r} w_{i,r}}} \sqrt{1-\mu^2 } \mu^n\nonumber\\
	&\times \left\{
	\e^{-i k_{s,z} L/2} \frac{u_n\left[w_{s,r}(k_{s,\perp}-k_0)\right]}{\sqrt{1/w_{s,r}}}\right\}\label{eq:radialdecomp}\end{align}

where $\mu$ is a singular value scaling factor while $w_s$ and $w_i$ are width parameters for the signal and the idler, respectively
\begin{align}
	|\mu|&=\frac{\sqrt{A_{ss}A_{ii}}}{|A_{si}|}-\sqrt{\frac{A_{ss}A_{ii}}{A_{si}^2}-1}\nonumber\\
	w_{s,r}&=\sqrt{2 A_{ss}}\left( 1-\frac{A_{si}^2}{A_{ss} A_{ii}} \right)^{1/4} &
	w_{i,r}&=\sqrt{2 A_{ii}}\left( 1-\frac{A_{si}^2}{A_{ss} A_{ii}} \right)^{1/4}.\label{eq:radialcoeff}
\end{align}
The coefficients of the quadratic form matrix $\tens{A}$ are to be taken from (\ref{eq:quadratic_nowalk-off}). As outlined in \eqref{eq:svd_separation}, once we have the SVDs of both $K(k_{i,\perp},k_{s,\perp})$ in (\ref{eq:radialdecomp}) and $\Phi(\varphi_i,\varphi_s)$ in (\ref{eq:angularexp}), we get the resulting $S_{is}$ decomposition for the no walk-off $\rho=0$ case
\begin{multline}
	S_{is}(k_{i,\perp},\varphi_i;k_{s,\perp},\varphi_s)=
	\sum_{n=0}^\infty \sum_{m=-\infty}^\infty
	\left\{\e^{i k_{i,z} L/2} \frac{u_n\left[w_{i,r}(k_{i,\perp}-k_0)\right]}{\sqrt{1/w_{i,r}}} \frac{\exp(-im\varphi_i)}{\sqrt{2\pi}}\right\}\\
\times
	\chi P L \frac{\sqrt{2\pi}\sqrt{1-\mu^2 }}{k_0 \sqrt{w_{s,r} w_{i,r}}}
	 \mu^n \exp\left(-\frac{m^2}{2k_0^2 w_p^2}\right)
\left\{\e^{-i k_{s,z} L/2} \frac{u_n\left[w_{s,r}(k_{s,\perp}-k_0)\right]}{\sqrt{1/w_{s,r}}} \frac{\exp(im\varphi_s)}{\sqrt{2\pi}}\right\}
.\label{eq:nowalk-offsolution}
\end{multline}

The resulting modes in a no walk-off setting are optical vortices. We discuss this result in Section 4.

\subsection{Significant walk-off}

The next case of our interest is the setting with a significant walk-off, i.e. when the total walk-off at the end of the crystal is of the order of the pump diameter $L \rho \approx w_p$. Again, we take the scattering kernel $S_{is}$ given in (\ref{eq:kernelexpquadratic}). The quadratic form $\tens{A}$ has nonzero mixed part, which prevents us from the separation of variables. In addition, it contains trigonometric functions of the sum of angles $\varphi_s+\varphi_i$.

Let us first consider the physical picture of the significant walk-off setting. An efficient amplification of a beam may be obtained when it travels along the pump. It happens when $\varphi_s+\varphi_i=0$ or $\varphi_s+\varphi_i=2\pi$ which corresponds to either signal or idler propagating along the pump. As the distinction between the signal and the idler is illusory, we may assume that the signal is propagating along the pump's walk-off, approximating all expressions containing $\varphi_s+\varphi_i$ around $0$. We will hold the quadratic term in $A_{\varphi\varphi}$ and only the constant term everywhere else in $\tens{A}$. Consequently, terms $A_{\varphi s}=A_{s \varphi}$ and $A_{\varphi i}=A_{i \varphi}$ disappear. Thus the quadratic form reads
\begin{align}
	\tens{A}=\left[
	\begin{array}{ccc}
	\frac{k_0^2 w_p^2}{2}+\frac{L^2}{20}k_0^2 \rho^2 ({\scriptstyle\frac{\varphi_s+\varphi_i}{2}})^2 &	 0 &	 0 \\
	 0 &
	 	 \frac{w_p^2}{2}+\frac{L^2}{20}(a_s-\rho)^2	&
	 -\frac{w_p^2}{2}+\frac{L^2}{20}(a_s-\rho)(a_i+\rho)\\
	 0 & 
	 -\frac{w_p^2}{2}+\frac{L^2}{20}(a_s-\rho)(a_i+\rho) &
	 \frac{w_p^2}{2}+\frac{L^2}{20}(a_i+\rho)^2
	\end{array}
	\right]\label{eq:kernelexpquadratic-walk-off}
\end{align}
Again, we have separated the variables $S_{is}=K(k_{i,\perp},k_{s,\perp}) \Phi(\varphi_i,\varphi_s)$ according to the scheme given in \eqref{eq:svd_separation}. The radial part $K(k_{i,\perp},k_{s,\perp})$ decomposes with the Mehler's formula \eqref{eq:hopropagator} as it did for the no-walk-off case. All the coefficients are the same as those given in (\ref{eq:radialcoeff}), with the quadratic form coefficient matrix $\tens{A}$ taken from a significant walk-off approximation matrix \eqref{eq:kernelexpquadratic-walk-off}.

The angular part $\Phi(\varphi_i,\varphi_s)$ contains not only $(\varphi_s+\varphi_i)^2$ but also $(\varphi_s+\varphi_i)^2(\varphi_s-\varphi_i)^2$ terms. They are difficult to handle, so we try the following approximation, with real variables p, q, x, and y
\begin{align}
	\exp\left[-p^2(1+q^2 y^2)x^2\right]\approx\exp\left(-p^2 x^2 -p^2q^2\langle x^2\rangle y^2\right)=\exp\left(-p^2 x^2 - \frac{1}{2}q^2 y^2\right)\label{eq:4to2}.
\end{align}
That is, instead of the $x^2y^2$ term, we took $y^2$ times mean $x^2$, averaged over the $\exp\left(-p^2 x^2\right)$ distribution. The rough approximation \eqref{eq:4to2} is numerically checked to produce similar characteristic modes and gain parameters as long as $q^2\ll 1$. In our case $q^2=L^2 \rho^2/(40w_p^2)$, so the total walk-off cannot be much larger than the pump width $L \rho \ll \sqrt{40} w_p$. With the approximation \eqref{eq:4to2} the angular part $\Phi(\varphi_i,\varphi_s)$ is a quadratic form of $\varphi_s$ and $\varphi_i$. Utilizing once again Mehler's formula \eqref{eq:hopropagator} we decompose $\Phi(\varphi_i,\varphi_s)$ in the basis of Hermite-Gaussian modes
\begin{align}
	\Phi(\varphi_i,\varphi_s)&=\exp \left[ -\frac{k_0^2 w_p^2}{2}(\varphi_s-\varphi_i)^2 -\frac{L^2 \rho^2}{20 w_p^2}\left(\frac{\varphi_s+\varphi_i}{2}\right)^2\right]\nonumber\\
	&=\sum_{m=0}^\infty  \left[ \frac{u_m(k_0 w_\varphi \varphi_i)}{\sqrt{1/(k_0 w_\varphi)}} \right]
	\frac{1}{k_0 w_\varphi} \sqrt{\pi} \sqrt{1-\mu_\varphi^2} \mu_\varphi^m \left[
	\frac{u_m(k_0 w_\varphi \varphi_s)}{\sqrt{1/(k_0 w_\varphi)}}\right]
\end{align}
where $\mu_\varphi$ is the singular value scaling factor while $k_0 w_\varphi$ is the angular width parameter for both the signal and the idler
\begin{align}
	\mu_{\varphi}&=\left(1-\frac{1}{2\sqrt{10}}\frac{L \rho}{k_0 w_p^2}\right)/\left(1+\frac{1}{2\sqrt{10}}\frac{L \rho}{k_0 w_p^2}\right) &
	k_0 w_\varphi&=\frac{1}{\sqrt[4]{10}}\sqrt{L k_0 \rho}.\label{eq:angularcoeff}
\end{align}
Hence, we obtain the SVD of the scattering kernel $S_{is}$ in the significant walk-off approximation \eqref{eq:kernelexpquadratic-walk-off} with separated variables \eqref{eq:svd_separation}, which reads
\begin{align}
	S_{is}(k_{i,\perp},\varphi_i;k_{s,\perp},\varphi_s)= \sum_{n=0}^\infty \sum_{m=0}^\infty
	\left\{\e^{i k_{i,z} L/2} \frac{u_n\left[w_{i,r}(k_{i,\perp}-k_0)\right]}{\sqrt{1/w_{i,r}}} \frac{u_m(k_0 w_\varphi \varphi_i)}{\sqrt{1/(k_0 w_\varphi)}}\right\}
	\nonumber\\
\times 
\chi P L \frac{w_p \sqrt{\pi} \sqrt{1-\mu^2 }  \sqrt{1-\mu_\varphi^2}}{k_0 w_\varphi \sqrt{w_{s,r} w_{i,r}}}  \mu^n \mu_\varphi^m
\left\{\e^{-i k_{s,z} L/2} \frac{u_n\left[w_{s,r}(k_{s,\perp}-k_0)\right]}{\sqrt{1/w_{s,r}}} \frac{u_m(k_0 w_\varphi \varphi_s)}{\sqrt{1/(k_0 w_\varphi)}}\right\}
.\label{eq:strongwalk-offsolution}
\end{align}
Further discussion of the modes is given in Section 4.

\section{Characteristic modes of the amplifier}

In Section 3, we applied a series of approximations to obtain the singular value decomposition of the scattering kernel $S_{is}$ \eqref{eq:pert_sis}. The SVDs for the no walk-off \eqref{eq:nowalk-offsolution} and significant walk-off \eqref{eq:strongwalk-offsolution} settings were compared with the general Bloch-Messiah reduction \eqref{eq:bloch-messiah-cs}. We identified the signal $\psi_{n,m}^{out}(k_{s,\perp},\varphi_s)=\psi_{n,m}^{in*}(k_{s,\perp},\varphi_s)$ and the idler $\phi_{n,m}^{out}(k_{i,\perp},\varphi_i)=\phi_{n,m}^{in*}(k_{i,\perp},\varphi_i)$ modes - see \eqref{eq:bloch-messiah-cs} and Fig. \ref{fig:bloch-messiah}. In particular, the highest achievable amplitude amplification $\xi_{0,0}$ is obtained with the seed $\psi_{0,0}^{in}(k_{s,\perp},\varphi_s)$  on the input, which produces $\psi_{0,0}^{in*}(k_{s,\perp},\varphi_s)$ on the output. It needs to be remembered that the amplitudes in the spatial frequency domain are effectively the far-field image.

\subsection{No walk-off}

\begin{figure}[htbp]
	\centering
	\begin{tabular}[t]{ccc}
		\includegraphics[width=0.30\textwidth]{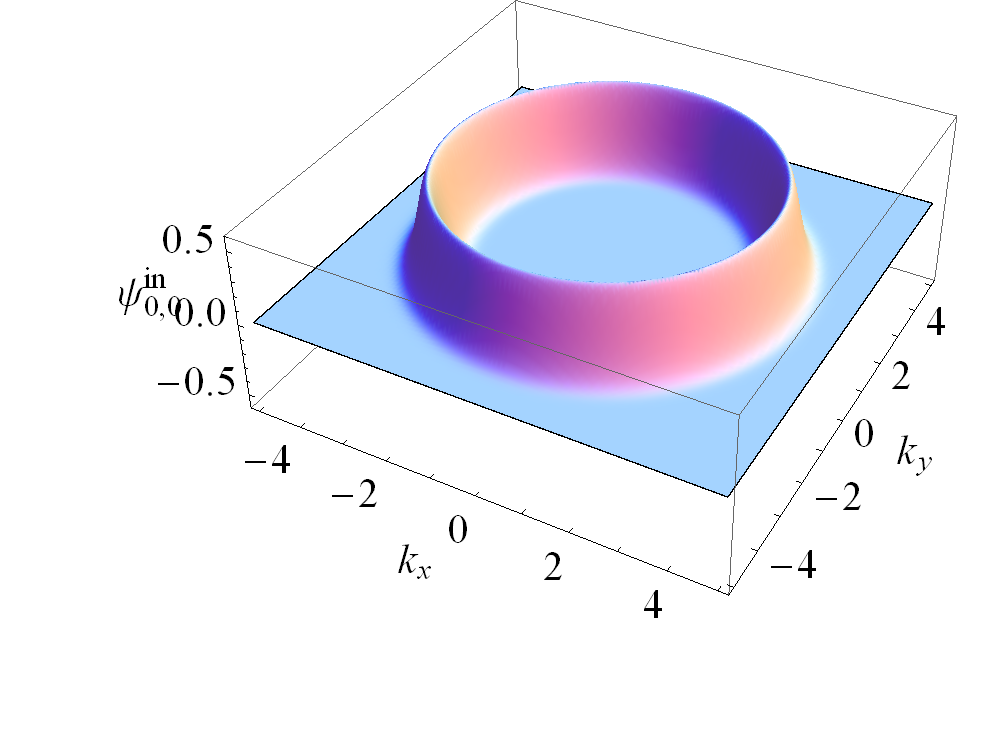}&
		\includegraphics[width=0.30\textwidth]{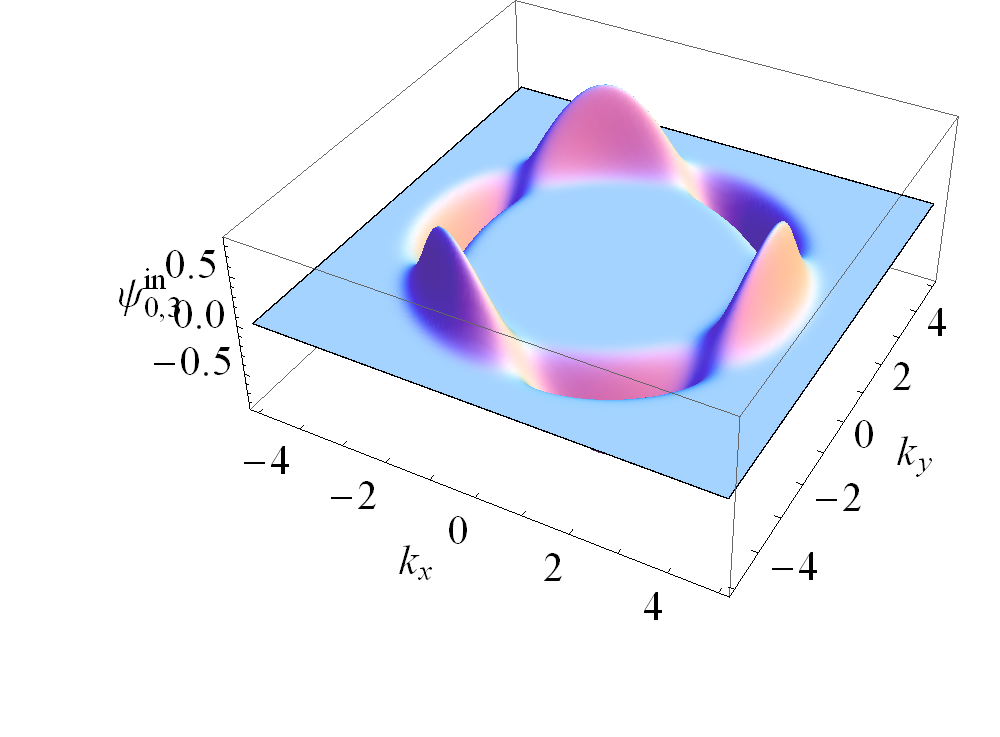}&
		\includegraphics[width=0.30\textwidth]{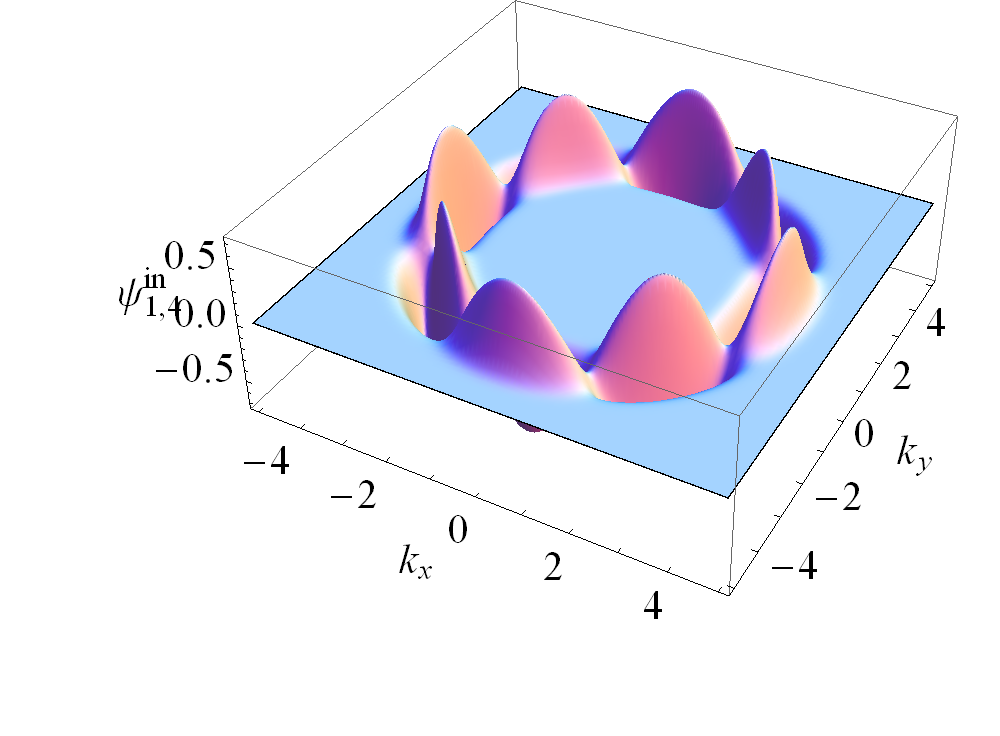}
				\end{tabular}
	\caption{Three different modes for the no walk-off case: $\psi_{0,0}^{in}(k_{s,\perp},\varphi_s)$, $\psi_{0,3}^{in}(k_{s,\perp},\varphi_s)$ and $\psi_{1,4}^{in}(k_{s,\perp},\varphi_s)$. The phase $\e^{-i k_{s,z} L/2}$ is not shown. The figure represents the real part of the modes which, due to degeneracy, are also proper characteristic modes.}
	\label{fig:modes_nowalkoff}
\end{figure}

In the setting with no walk-off $\rho = 0$ the modes resulting from the solution \eqref{eq:nowalk-offsolution} are optical vortices
\begin{align}
	\psi_{n,m}^{out*}(k_{s,\perp},-\varphi_s)=\psi_{n,m}^{in}(k_{s,\perp},\varphi_s)&= \e^{-i k_{s,z} L/2}
	\frac{u_n\left[w_{s,r}(k_{s,\perp}-k_0)\right]}{\sqrt{1/w_{s,r}}} \frac{\exp(im\varphi_s)}{\sqrt{2\pi}}\nonumber\\
	\phi_{n,m}^{out*}(k_{i,\perp},-\varphi_i)=\phi_{n,m}^{in}(k_{i,\perp},\varphi_i)&= \e^{-i k_{i,z} L/2}
	\frac{u_n\left[w_{i,r}(k_{i,\perp}-k_0)\right]}{\sqrt{1/w_{i,r}}} \frac{\exp(-im\varphi_i)}{\sqrt{2\pi}}.
\end{align}
The minus sign at the $\varphi_s$ and $\varphi_i$ angles is a nontrivial consequence of twofold degeneracy of the singular values, which changes relations \eqref{eq:symmetry}.
The above modes are trigonometric functions on the down-conversion cone as plotted in Fig. \ref{fig:modes_nowalkoff}. Their shape in the radial direction is that of a Hermite-Gaussian function with the width parameter
\begin{align}
	w_{s,r} &= \sqrt[4]{1-\nu^{-1}}\sqrt{w_p^2+\frac{L^2}{10} a_s^2} &
	w_{i,r} &= \sqrt[4]{1-\nu^{-1}}\sqrt{w_p^2+\frac{L^2}{10} a_i^2}\label{eq:nowalk-offmodes},
\end{align}
for the signal and the idler respectively, where $\nu=\left(w_p^2+\frac{L^2}{10} a_s^2 \right)\left( w_p^2+\frac{L^2}{10}a_i^2 \right)/\left( -w_p^2+\frac{L^2}{10}a_s a_i \right)^2$. The gain parameters are
\begin{align}
		\xi_{n,m} &=  \chi P L \frac{\sqrt{2 \pi}\sqrt{1-\mu^2 }}{k_0 \sqrt{w_{s,r} w_{i,r}}}
	 \mu^n \exp\left(-\frac{m^2}{2k_0^2 w_p^2}\right), &
	 	|\mu| &= \sqrt{\nu}-\sqrt{\nu-1}.
\end{align}
The characteristic modes are optical vortices \cite{AllenPRA92}, thanks to the $\exp(i m \varphi_s)$ factor. Each of them carry orbital angular momentum, $\hbar m$ per photon. As the spontaneous down-conversion creates entangled photon pairs with the opposite angular momenta \cite{TorresJoOBQaSO05}, they may be used in the field of quantum cryptography \cite{MairN01}. The advantages of using entangled vortex states as media for quantum information include easy measurement \cite{LeachPRL02} and the possibility to create Hilbert space of an arbitrary dimension \cite{TorresPRA03}.

To obtain the above decomposition we assumed that: ${1}/{w_p}\ll k_0$, ${1}/{w_{s,r}}\ll k_0$ and ${1}/{w_{i,r}}\ll k_0$. This is approximately equivalent to the requirement that beam diffraction angles are much smaller than the angle between signal or idler and the pump. 

\subsection{Significant walk-off}

\begin{figure}[htbp]
	\centering
	\begin{tabular}[t]{ccc}
		\includegraphics[width=0.30\textwidth]{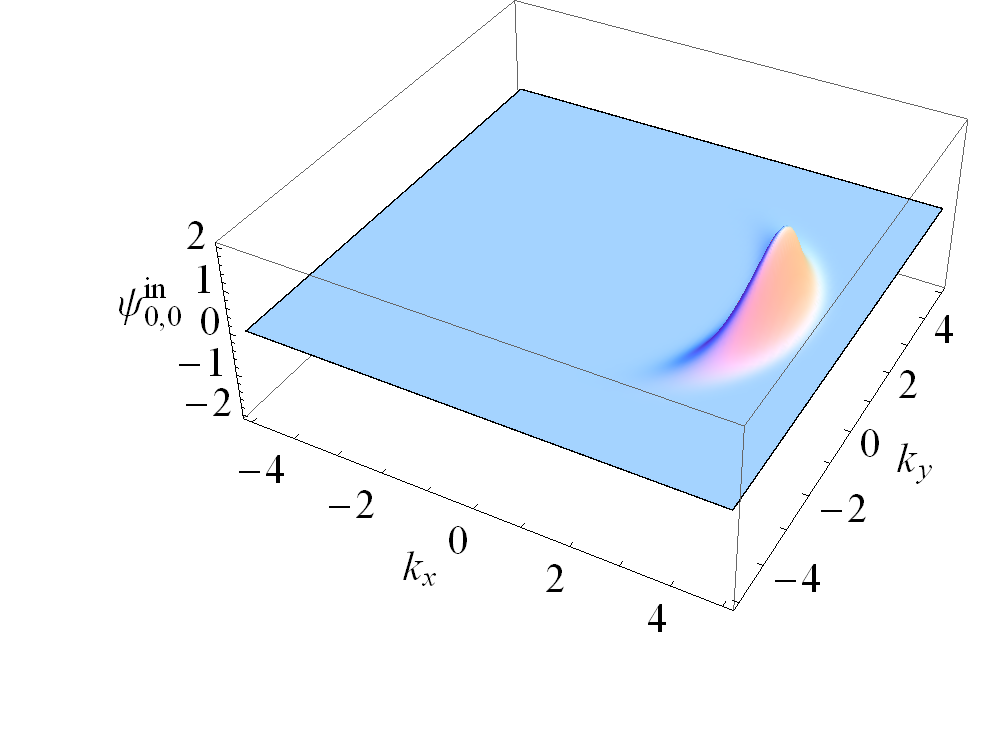}&
		\includegraphics[width=0.30\textwidth]{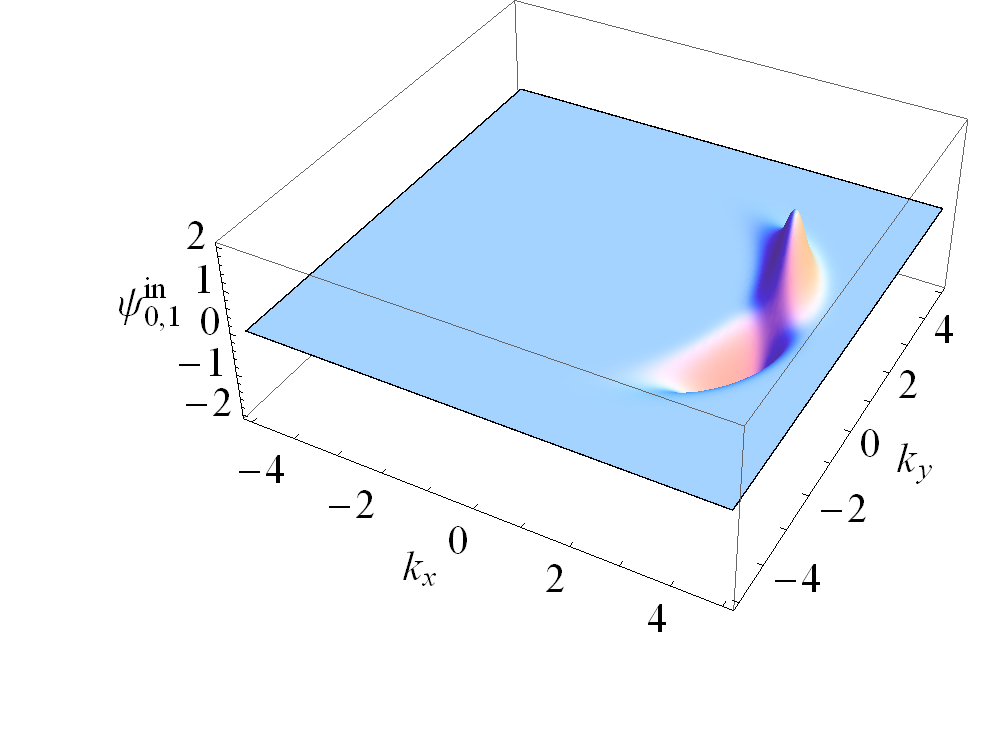}&
		\includegraphics[width=0.30\textwidth]{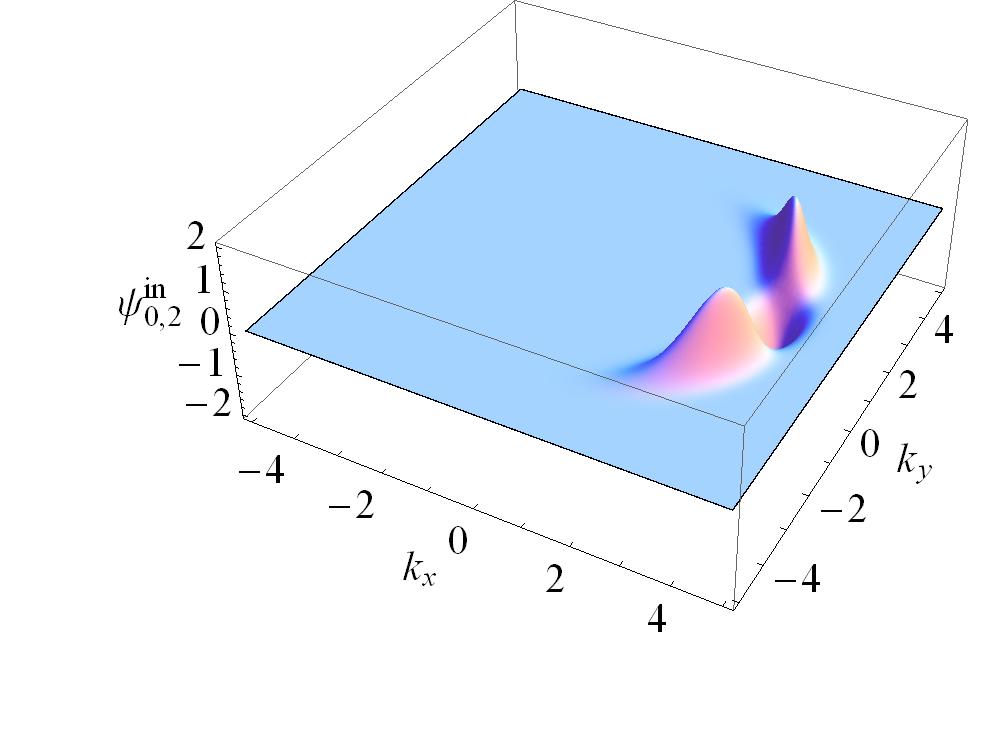}
				\end{tabular}
	\caption{First three modes for the significant walk-off case: $\psi_{0,0}^{in}(k_{s,\perp},\varphi_s)$, $\psi_{0,1}^{in}(k_{s,\perp},\varphi_s)$ and $\psi_{0,3}^{in}(k_{s,\perp},\varphi_s)$. The phase $\e^{-i k_{s,z} L/2}$ is not shown. The modes are bent elliptic Gaussian-Hermite beams. When angular width is sufficiently small, the mode with the highest amplification $\psi_{0,0}^{in}(k_{s,\perp},\varphi_s)$ is just an elliptic Gaussian.}
	\label{fig:modes_strongwalkoff}
\end{figure}

In the setting with a significant walk-off $\rho > 0$ the modes resulting from \eqref{eq:strongwalk-offsolution} are elliptic Hermite-Gaussian beams
\begin{align}
	\psi_{n,m}^{out*}(k_{s,\perp},\varphi_s)=\psi_{n,m}^{in}(k_{s,\perp},\varphi_s)&=
	\e^{-i k_{s,z} L/2}
	\frac{u_n\left[w_{s,r}(k_{s,\perp}-k_0)\right]}{\sqrt{1/w_{s,r}}}
	\frac{u_m(k_0 w_\varphi \varphi_s)}{\sqrt{1/(k_0 w_\varphi)}}\nonumber\\
	\phi_{n,m}^{out*}(k_{i,\perp},\varphi_i)=\phi_{n,m}^{in}(k_{i,\perp},\varphi_i)&=
	\e^{-i k_{i,z} L/2}
	\frac{u_n\left[w_{i,r}(k_{i,\perp}-k_0)\right]}{\sqrt{1/w_{i,r}}}
	\frac{u_m(k_0 w_\varphi \varphi_i)}{\sqrt{1/(k_0 w_\varphi)}}.\label{eq:strongwalkoffmodes}
\end{align}
as plotted in Fig. \ref{fig:modes_strongwalkoff}. That is, the characteristic modes are 2D Hermite-Gaussian functions, curved on the down-conversion cone. Their width parameters are
\begin{align}
	w_{s,r} &= \sqrt[4]{1-\nu^{-1}}\sqrt{w_p^2+\frac{L^2}{10} (a_s-\rho)^2} &
	w_{i,r} &= \sqrt[4]{1-\nu^{-1}}\sqrt{w_p^2+\frac{L^2}{10} (a_i+\rho)^2}\nonumber\\
	k_0 w_\varphi &=\frac{1}{\sqrt[4]{10}}\sqrt{L k_0 \rho} &
\nu&=\frac{\left[w_p^2+\frac{L^2}{10} (a_s-\rho)^2 \right]\left[ w_p^2+\frac{L^2}{10}(a_i+\rho)^2 \right]}{\left[ -w_p^2+\frac{L^2}{10}(a_s-\rho) (a_i+\rho) \right]^2}.
\label{eq:swowidths}
\end{align}
The respective gain parameters are
\begin{align}
	\xi_{n,m} &=
	\chi P L \frac{w_p \sqrt{\pi} \sqrt{1-\mu^2 }  \sqrt{1-\mu_\varphi^2}}{k_0 w_\varphi \sqrt{ w_{s,r} w_{i,r}}}  \mu^n \mu_\varphi^m, &
		|\mu| &= \sqrt{\nu}-\sqrt{\nu-1}, &
		\mu_{\varphi}&=\frac{1-\frac{1}{2\sqrt{10}}\frac{L \rho}{k_0 w_p^2}}{1+\frac{1}{2\sqrt{10}}\frac{L \rho}{k_0 w_p^2}}.\label{eq:swomu}
\end{align}
To obtain the above decomposition we assumed that: $1/w_p\ll k_0$, $1/w_{s,r}\ll k_0$, $1/w_{i,r}\ll k_0$, $1/k_0 w_\varphi\ll \pi$ and $L\rho/\sqrt{40}\ll w_p$. In other words, beam diffraction angles need to be much smaller than the phase matching angles, the total walk-off cannot be much larger than the pump waist and the mode arc width has to be smaller than $\pi$.

When angular width is relatively small, the modes are just elliptic Hermite-Gaussian functions \cite{WalbornPRA05} in spatial frequencies. The explicit condition is $w_{s,r} \ll k_0 w_\varphi^2$ for the signal and $w_{i,r} \ll k_0 w_\varphi^2$ for the idler. Then the Fourier transform of \eqref{eq:strongwalkoffmodes} yields in the spatial mode functions in the near field
\begin{align}
	\psi_{n,m}^{out*}(-x,-y)=\psi_{n,m}^{in}(x,y)&=
	\e^{-i k_0 x}
	\frac{u_n(\frac{x+a_s L/2}{w_{s,r}})}{\sqrt{w_{s,r}}}
	\frac{u_m(\frac{y}{w_\varphi})}{\sqrt{w_\varphi}}\nonumber\\
	\phi_{n,m}^{out*}(-x,-y)=\phi_{n,m}^{in}(x,y)&=
	\e^{i k_0 x}
	\frac{u_n(\frac{x-a_i L/2}{w_{i,r}})}{\sqrt{w_{i,r}}}
	\frac{u_m(\frac{y}{w_\varphi})}{\sqrt{w_\varphi}}\label{eq:strongwalkoffmodesspace}
\end{align}
In particular the $\psi_{0,0}^{in}(x,y)$ is an elliptical Gaussian beam, traveling along the ideal phase matching cone, that is, at the angle $\arctan(a_s)$ and passing through the center of the crystal. Its widths are $w_{s,r}$ in the radial direction and $w_\varphi$ in the angular direction. The mode is optimal in the terms of maximal achievable gain, as well as the signal-to-noise ratio.

\section{Numerical simulations}

Above, we have derived the singular value decompositions \eqref{eq:nowalk-offsolution} and \eqref{eq:strongwalk-offsolution} from the low gain regime scattering kernel $S_{is}$ given in \eqref{eq:pert_sis}. However, along with mathematically well justified approximations we have taken a few less rigorous steps, especially (\ref{eq:sinc5}) and (\ref{eq:4to2}). Furthermore, the separation of variables \eqref{eq:svd_separation} needs to be verified, as well as the Taylor series approximation around $\varphi_s+\varphi_i=0$ \eqref{eq:kernelexpquadratic-walk-off}. To confirm the validity of the derivation of the characteristic modes, we performed a numerical simulation.

For each crucial step in the approximation of the scattering kernel $S_{is}$, we performed its numerical SVD. Technically, we calculated a discretized kernel in a cylindrical coordinate system \eqref{eq:cylindrical}, corrected with the proper Jacobian. It was too memory consuming to calculate $S_{is}$ over the entire rectangular sector of the coordinate grid, so we calculated its values only in those regions which have a potential to contribute significantly, and we employed sparse arrays. 
Since $S_{is}$ \eqref{eq:pert_sis} has the pump amplitude \eqref{eq:pump} as a factor, we took into account only those regions in ($\vec{k}_{i,\perp}$, $\vec{k}_{s,\perp}$) space for which the values of $A_p(\vec{k}_{s,\perp}+\vec{k}_{i,\perp})$ are significant. In the cylindrical coordinates this happens when
\begin{align}
	|k_{s,\perp}-k_{i,\perp}|&<\frac{2.5}{w_p} & &\text{and} &
	|\varphi_s-\varphi_i|&<\frac{2.5}{k_0 w_p}.\label{eq:pumprest}
\end{align}
We chose the constant $2.5$ as it covers over $0.999$ of the pump intensity. Moreover, we restricted ourselves to values close to the ideal phase matching setting $k_{s,\perp}$, $k_{i,\perp}\simeq k_0$ and $\varphi_s$, $\varphi_i\simeq 0$. The extent of the grid must be much greater than the zeroth mode, while the mesh must be much smaller than the peak sizes. The requirement is especially important for the $\sinc$ peak, as a too low resolution may spoil its oscillating shape. To check if the simulation works properly, we compared results for the same physical data but plotted over a twice finer grid or at a twice broader range. 

\begin{figure}[htbp]
	\centering
	\begin{tabular}[t]{cc}
		\includegraphics[width=0.45\textwidth]{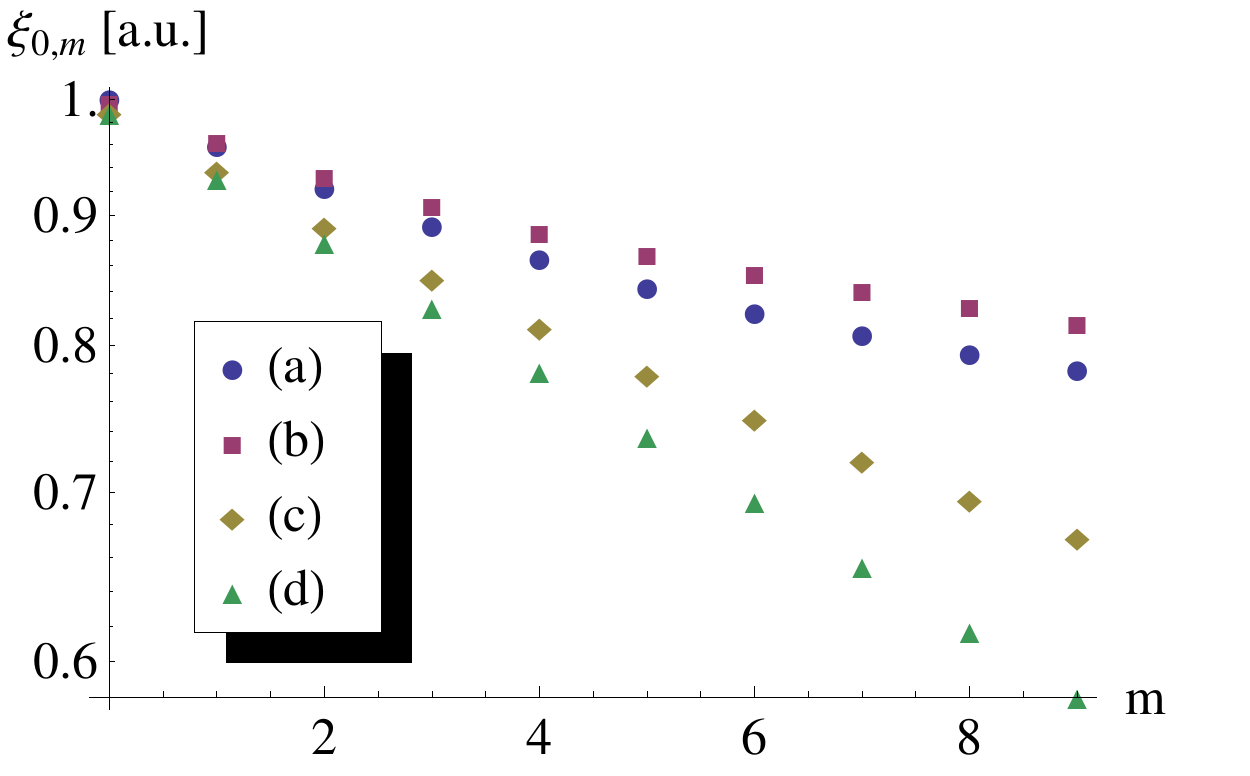}&
	  \includegraphics[width=0.45\textwidth]{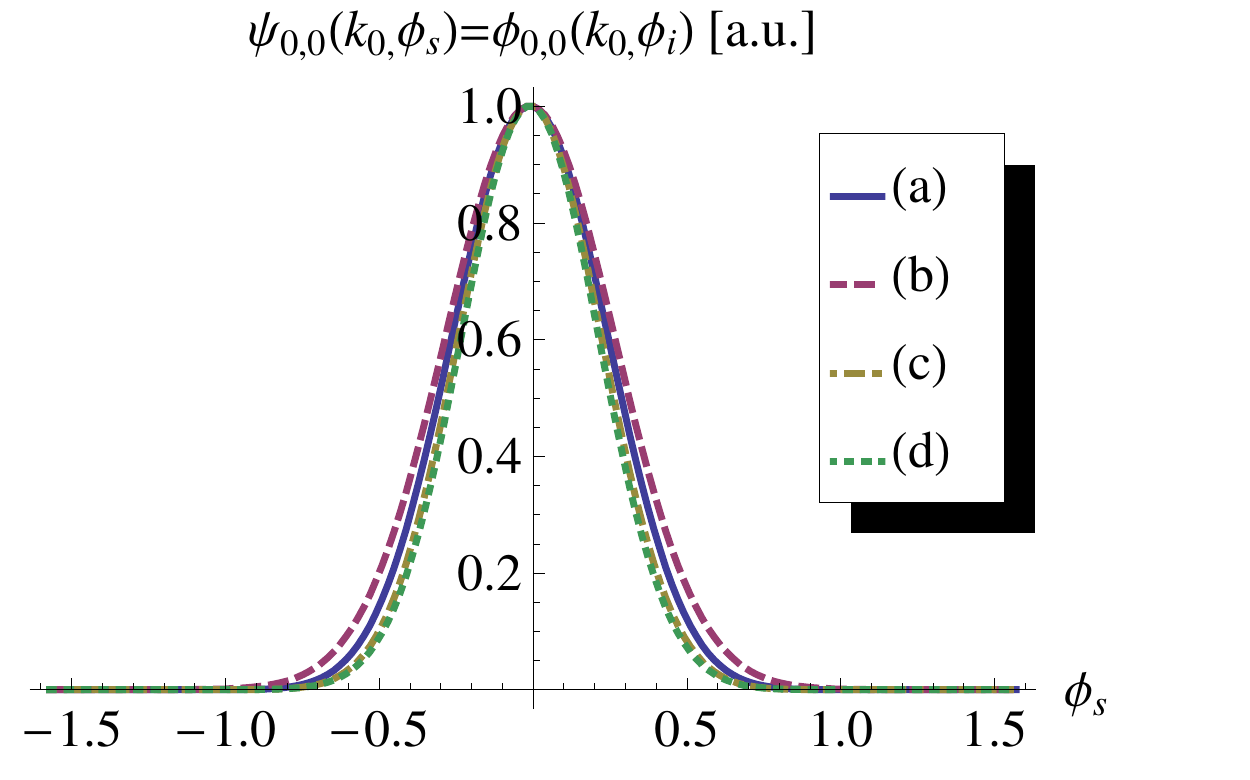}\\
	  i) & ii)\\
		\includegraphics[width=0.45\textwidth]{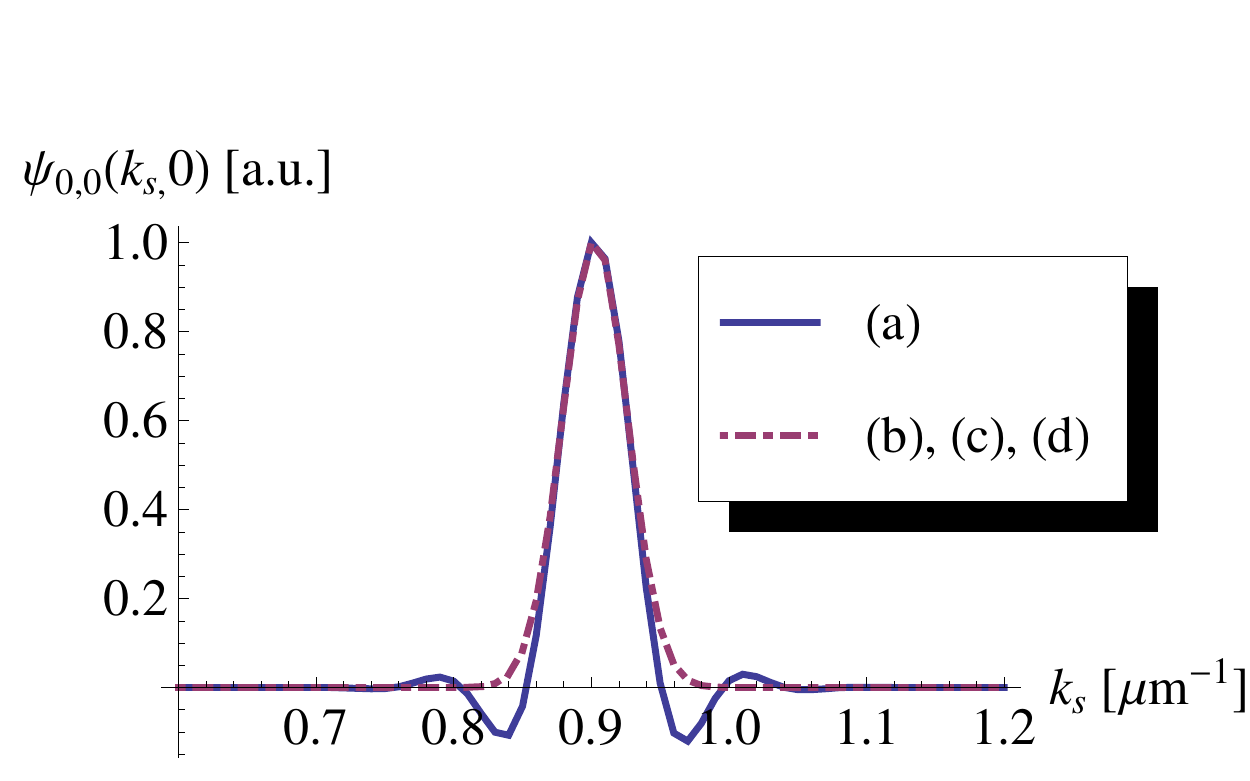}&
		\includegraphics[width=0.45\textwidth]{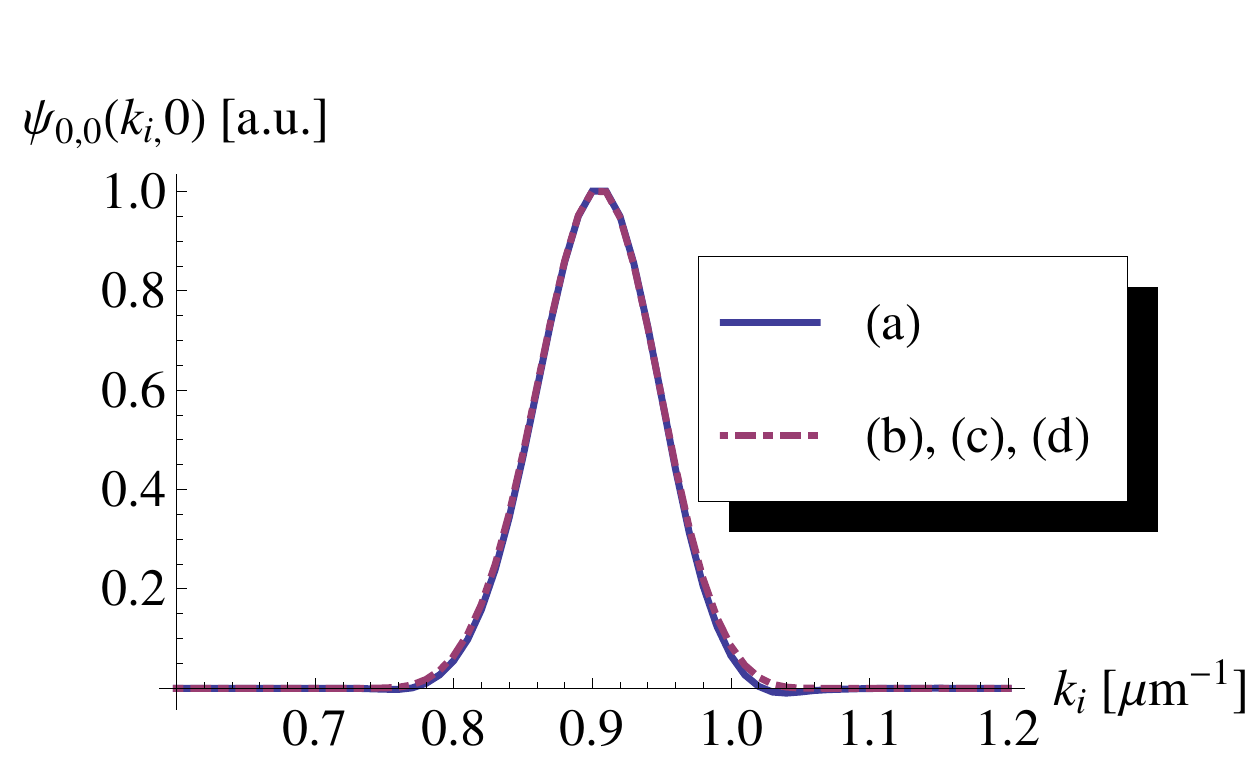}\\
		iii) & iv)
				\end{tabular}
	\caption{Comparison of i) the gain parameters $\xi_{0,m}$ and ii-iv) sections though mode functions obtained for four different approximations of the scattering kernel $S_{is}$: (a) the original with $\sinc$ \eqref{eq:pert_sis}, (b) the kernel with $\exp$ \eqref{eq:kernelexp}, (b') the quadratic form approximation \eqref{eq:kernelexpquadratic}, (c) the kernel with the mixed term removed \eqref{eq:kernelexpquadratic-walk-off} and (d) the final result \eqref{eq:strongwalk-offsolution}. As (b) and (b') were indistinguishable, they are represented only by (b). 
Calculations were carried out for a $\beta-$BBO crystal with $L=1\hbox{mm}$, $\theta=30^\circ$, $w_p=20\hbox{$\mu$m}$, $\lambda_p=0.4\hbox{$\mu$m}$, $\lambda_s=0.6\hbox{$\mu$m}$. This corresponds to significant walk-off, $\rho L/w_p\approx 2.4$. The kernels have been calculated on a grid of $61\times 161$ points, covering the range of $[0.6\hbox{$\mu$m}^{-1},1.2\hbox{$\mu$m}^{-1}]\times[-\frac{\pi}{2},\frac{\pi}{2}]$.}
	\label{fig:numerical}
\end{figure}

After verifying the reliability of numerical SVD, we have compared results from different scattering kernel $S_{is}$ approximations for a significant walk-off setting. The results are illustrated in Fig. \ref{fig:numerical}.
For this data the zeroth singular value $\xi_{0,0}$ was preserved by the approximation, within a $2\%$ margin of error. However, other gain parameters changed visibly with every approximation. The radial shape of the zeroth mode was altered only by the sinc-exp approximation \eqref{eq:sinc5}. Even though the consecutive approximations modified the angular width $w_{s,\varphi}$, the modes were still qualitatively Hermite-Gaussians. Ten first characteristic modes were verified to be indeed well separable. They could be reproduced by the product of radial and angular functions in over $98\%$ of their intensity. 

\section{Practical signal optimization and noise reduction}

In the previous chapters we worked in the perturbative regime, that is, we assumed that the gain parameters are small $\xi_{n,m}\ll 1$. In particular, we found that the gain parameters are proportional to the pump amplitude $\xi_{n,m}\propto P$ and that the characteristic modes do not change with the pump amplitude $P$. However, a useful amplifier requires gain parameters much higher than $1$. Hence, we need to extrapolate the results. We conjuncture that the proportionality relation $\xi_{n,m}\propto P$ holds also for higher gains, while the mode functions remain unchanged. This is strictly true for a 1-D parametric amplifier we considered in Section 2 and was numerically verified for a waveguide amplifier pumped with ultrashort pulses \cite{WasilewskiPRA06a}. 

In a seeded OPA, two processes occur in parallel: amplification of the seed beam and generation of the spontaneous parametric fluorescence. They do not influence one another as long as the saturation effects can be neglected. Both of these processes can be described within the framework of the developed model. The latter process typically sets the noise level of the amplifier. Since the total amount of fluorescence depends only on the pump parameters and not on the seed, it is best to chose the seed beam size which provides maximum possible amplification. This is exactly the beam described by the fundamental input mode $\psi^{in}_{0,0}(\vec{k}_{s,\perp})$ given in \eqref{eq:strongwalkoffmodes} and \eqref{eq:strongwalkoffmodesspace}. In a typical situation, this is an elliptic Gaussian beam of size $w_{s,r}\times w_{\varphi}$ propagating at an angle $\arctan(a_s)$ inside a nonlinear crystal. In the interaction with the pump light the seed beam is amplified by a factor $G=\exp(2\xi_{0,0})$ and a beam in a mode $\psi^{out}_{0,0}(\vec{k}_{s,\perp})$ is produced at the output.

An elliptical Gaussian beam can be easily converted to a Gaussian beam with the help of astigmatic optics. If astigmatic shaping of the input beam is unfeasible, one can use round Gaussian beam of the waist $w_s = \sqrt{w_{s,r} w_{\varphi}}$. In such case many input modes $\psi^{in}_{2n,2m}(\vec{k}_{s,\perp})$ will be excited, each of them will be amplified by a factor $\exp(2\xi_{2n,2m})$ and a distorted beam will be produced on the output. Typically we may neglect  higher order modes, since they both have lower gains and are only slightly excited. The most important effect of using a round input beam is reduced coupling of the seed light to the fundamental mode. This is described by the factor
\begin{align}
	\eta = \frac{4 w_{s,r} w_{s,\varphi}}{(w_{s,r}+ w_{s,\varphi})^2}\label{eq:eta}
\end{align}
which effectively reduces gain to $\eta G$.

Let us now proceed with the calculation of the intensity of the fluorescence noise emitted per unit time in a certain direction. From basic quantum-mechanical considerations we know that the number of photons scattered into mode $\psi^{out}_{n,m}(\vec{k}_{s,\perp})$ at a frequency of $\omega_s$ is equal to $\sinh^2 \xi_{n,m}\simeq \exp(2 \xi_{n,m})$. When we observe the output of the amplifier through an aperture, the contributions of distinct modes should be added incoherently. Thus, the spectral density of the number of photons is equal to the sum
\begin{align}
\langle\tilde n(\omega_s)\rangle=\sum_{n,m} \exp(2\xi_{n,m}) T_{n,m}\label{eq:transmission}
\end{align}
where $T_{n,m}$ is the transmission of the output mode $\psi^{out}_{n,m}(\vec{k}_{s,\perp})$ trough the observation aperture. 

To calculate the number of fluorescence photons emitted per unit time $d\langle n\rangle/dt$ we add
contributions from all the frequencies 
\begin{align}
	\frac{d\langle n\rangle}{dt}= \int d\omega_s \, \langle \tilde{n}(\omega_s)\rangle,\label{eq:totalnoise}
\end{align}
where the integral should be performed over the observation bandwidth. With such a formulation, the pump intensity and the number of fluorescence photons emitted per unit time $d\langle n\rangle/dt$ can become slowly varying time dependent quantities. However, let us focus on the simplest way of minimizing the total noise represented by the integral \eqref{eq:totalnoise}, that is minimizing
the noise power at each frequency $\tilde n(\omega_s)$ independently. We consider achieving this goal by spatial filtering of the output of the amplifier with an aperture.

\begin{figure}[htbp]
	\centering
	\includegraphics[width=0.50\textwidth]{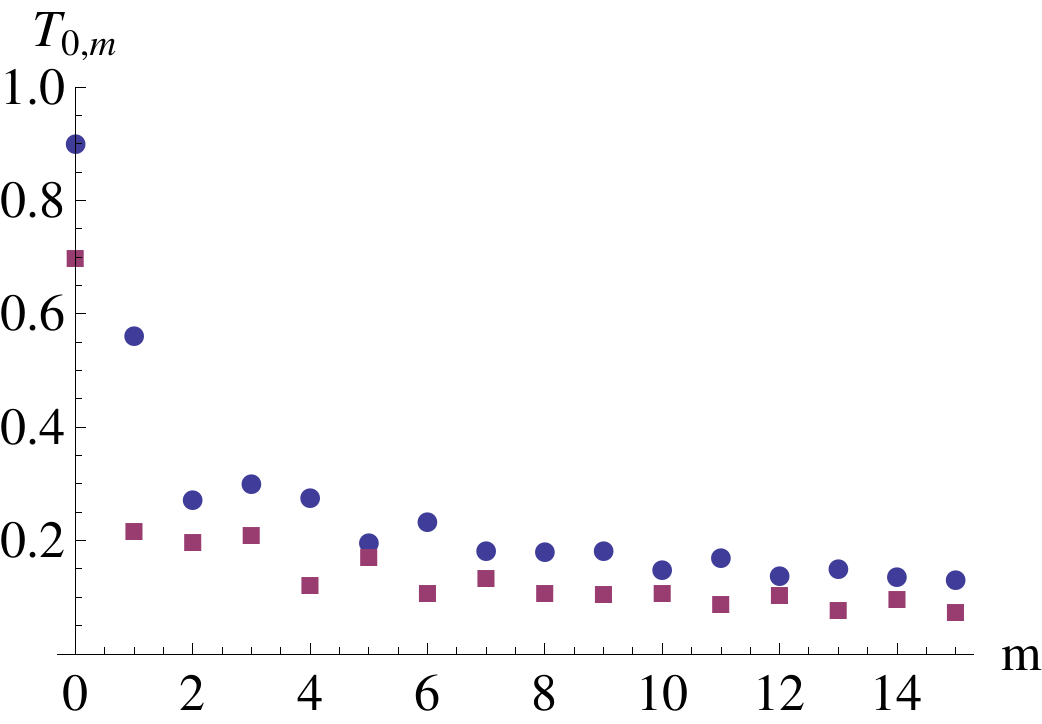}
	\caption{ Plot of intensity transmittance $T_{0,m}$ of the first 15 Hermite-Gaussian modes trough a slit set to transmit $70\%$ (squares) or $90\%$ (dots) of the zeroth mode.}
	\label{fig:mismatchaperture}
\end{figure}

As one my find in \eqref{eq:swomu} the gain parameters $\xi_{n,m}$ form geometric series with respect to both $n$ and $m$ as $\xi_{n,m}\propto \mu^n \mu_\varphi^m$. For typical data, $\mu$ is small and modes with $n>0$ can be neglected. On the other hand, $\mu_\varphi$ is often close to 1 and spatial filtering in the angular direction can become necessary. This can be accomplished by imaging the crystal onto a suitably oriented slit. One may then achieve high transmission for the fundamental mode $T_{0,0}$ carrying the useful signal, and significant suppression for the higher order mode $T_{0,m}$, as plotted in Fig. \ref{fig:mismatchaperture}. Results for arbitrary slit width can be checked with a script we provide online \cite{MigdalDownconversion}. The explicit expression for the intensity transmission of $\psi^{out}_{n,m}(\vec{k}_{s,\perp})$ mode through the slit of width $h$ is given by the integral
\begin{align}
	T_{n,m}=\int_{-h/2}^{h/2} dy\, u_m^2\left(\frac{y}{w_{\varphi}}\right),
\end{align}
where $u_m(y/w_\varphi)$ are the angular components of the mode functions \eqref{eq:strongwalkoffmodesspace}. For an aperture transmitting $70\%$ of the zeroth mode, the signal-to-noise ratio may be elevated up to $3.3$ times.

%The signal-to-noise ratio, or $\exp(2\xi_{0,0}) T_{0,0}/\langle\tilde n\rangle$ as in \eqref{eq:transmission}, can be optimized in two independent ways - manipulation with the gain parameters $\xi_{n,m}$ and the aperture transmittance.

Another approach to reducing fluorescence contained in high order modes is reducing angular gain decrement $\mu_\varphi$. Our simplified model predicts that it would approach zero when the following relation holds between the crystal length and the pump beam waist
\begin{align}
w_p=\frac{1}{\sqrt{2\sqrt{10}}} \sqrt{\frac{L\rho}{k_0}}.
%	2\sqrt{10}\, w_p^2 k_0=L \rho.
\label{eq:optimalpumpwidth}
\end{align}
Naturally the above relation should be considered approximate, because it relies on the numerical factor used in sinc approximation \eqref{eq:sinc5}. Fortunately, even a rough fit to the (\ref{eq:optimalpumpwidth}) should suffice.

\section{Conclusion}

We have developed a simple analytical model of a noncollinear parametric amplifier pumped with a focused monochromatic beam and utilizing type I phase matching. 
We found an approximate Bloch-Messiah reduction for a low-gain parametric amplifier for an arbitrary pair of signal and idler frequencies. The final result was expressed in a closed form in two special cases: zero walk-off of the pump beam or total walk-off of the order of the beam size. Characteristic modes of the first setting are optical vortices. The latter case corresponds to the typical experimental situation and we find that the characteristic modes are elliptic Hermite-Gaussian beams. We checked the validity of the approximations assumed during the derivation by comparing the analytical result to the numerical calculations for a typical range of parameters. Finally, we scaled the results of the reduction to a high gain regime using the results of our previous work \cite{WasilewskiPRA06a} and we discussed how to calculate the gain and fluorescence intensity.

In particular, we have calculated the fundamental mode of the amplifier, which has turned out to be elliptic Gaussian. Seeding the amplifier with a beam matching this mode yields the highest possible amplification. The output of the parametric amplifier contains both the amplified seed beam and optical noise due to parametric fluorescence. The Bloch-Messiah reduction allows us to directly calculate the amount of parametric fluorescence emitted by the amplifier. Further analysis shows that it can be reduced by either suitable spatial filtering of high-order modes or adjusting the pump beam diameter so that that those modes are suppressed. 

The results of our approximated model will never be as accurate as numerical simulations based on tracing the evolution of quasi-probability distributions \cite{ChwedenczukPRA08b}, but they can be instantly calculated for various configurations of parametric amplifiers using the Mathematica 6 script we provide online \cite{MigdalDownconversion}, and thus may provide valuable insight in the applications.

Our results may be also used as a starting point for numerical code modeling OPCPA with pump depletion. This would be accomplished by dividing the amplifier longitudinally into two parts: the front one, in which a gain of $10^3$--$10^6$ is reached without depleting the pump, and the rear one, where saturation occurs. In the first part, quantum phenomena play a role, while in the second part spontaneous fluorescence can be treated as classical noise. The front part can be treated with our model. This way the optimal input seed beam shape and spatially resolved intensity of the parametric fluorescence can be obtained.  The latter can serve as an initial condition for the numerical code solving the classical evolution of the fields in the rear part of the amplifier.

The model developed in this paper may be also helpful for optimizing photon pair sources, since it represents an alternative approach to the problem of finding spatial modes of the parametric fluorescence and optimal fiber coupling \cite{DraganPRA04}.

\subsection*{Acknowledgments}
We acknowledge insightful discussions with Konrad Banaszek and Czes{\l}aw Radzewicz,
as well as financial support from the Polish Government in the form of a scientific grant (2007-2009).

%\bibliography{downconversion2d}

\end{document}